\documentclass[format=acmsmall, review=false, screen=true]{acmart}

\PassOptionsToPackage{hyphens}{url}
\usepackage{hyperref}
\usepackage{url}

\usepackage{booktabs} 

\acmJournal{FACMP}
\acmVolume{0}
\acmNumber{1}
\acmArticle{1}
\acmMonth{2}

\setcopyright{acmcopyright}

\acmDOI{0000001.0000001}

\usepackage{array}
\newcolumntype{P}[1]{>{\centering\arraybackslash}p{#1}}
\newcolumntype{M}[1]{>{\centering\arraybackslash}m{#1}}
\newcolumntype{R}[1]{>{\arraybackslash}m{#1}}

\definecolor{orange}{rgb}{1,0.5,0}
\definecolor{graynode}{RGB}{20,20,20}
\definecolor{crimsonred}{RGB}{220,20,60}
\definecolor{darkgraynode}{gray}{0.5}
\definecolor{lightgraynode}{gray}{0.8}

\usepackage{rotate}
\usepackage{capt-of}
\usepackage{setspace}
\usepackage{mathtools}
\usepackage{pifont}

\definecolor{gray}{RGB}{20,20,20}
\definecolor{gray}{RGB}{0.7,0.7,0.7}
\definecolor{greencm}{RGB}{0,153,0}

\usepackage{tabularx}
\usepackage{booktabs}

\usepackage{comment}
\usepackage{paralist}
\setcopyright{none}
\usepackage{nicefrac}
\definecolor{thelightblue}{RGB}{0,191,255}

\usepackage{nicefrac}

\usepackage{mathtools}
\usepackage{blkarray, bigstrut} 

\usepackage{tikz}
\usepackage{verbatim}
\usetikzlibrary{arrows}
\usetikzlibrary{shapes,snakes}
\usetikzlibrary{decorations.pathmorphing} 
\usetikzlibrary{fit}					
\usetikzlibrary{backgrounds}	

\makeatletter
\global\let\tikz@ensure@dollar@catcode=\relax
\makeatother

\usepackage{subfigure}

\definecolor{thelightblue}{RGB}{0,191,255}
\definecolor{theblue}{RGB}{0,0,180}

\usepackage{blkarray}
\usepackage{bigstrut, booktabs}

\makeatletter
\renewcommand*\env@matrix[1][*\c@MaxMatrixCols c]{
\hskip -\arraycolsep
\let\@ifnextchar\new@ifnextchar
\array{#1}}
\makeatother

\usepackage{booktabs} 
\usepackage{enumitem}
\usepackage{subfigure}
\usepackage{rotating}
\usepackage{multirow}

\definecolor{mydarkblue}{RGB}{0, 20, 159} 
\definecolor{mydarkblue}{rgb}{0,0.08,0.45} 
\usepackage{hyperref}
\hypersetup{colorlinks=true,urlcolor=mydarkblue, 
citecolor=mydarkblue,  linkcolor=mydarkblue}

\DeclareSymbolFont{cmbrightop}{OT1}{cmbr}{m}{n}
\DeclareMathSymbol{\sfPsi}{\mathalpha}{cmbrightop}{9}
\let\hat\widehat

\usepackage{tikz}
\usepackage{verbatim}
\usetikzlibrary{arrows}
\usetikzlibrary{shapes,snakes}
\usetikzlibrary{decorations.pathmorphing} 
\usetikzlibrary{fit}					
\usetikzlibrary{backgrounds}	

\definecolor{gray}{RGB}{150,150,150}
\definecolor{theblue}{RGB}{0, 20, 159} 

\definecolor{myyellow}{RGB}{255,255,204}
\definecolor{myred}{RGB}{255,204,204}
\definecolor{myblue}{RGB}{0,200,255}
\definecolor{mygreen}{RGB}{80,220,80}

\newcommand{\eg}{\emph{e.g.}}
\newcommand{\ie}{\emph{i.e.}}

\newcommand{\wrt}{\emph{w.r.t.}\ }

\newtheorem{Property}{\bfseries{Property}}

\newtheorem{Definition}{\hspace{-1em}\bfseries{Definition}}
\newtheorem{Claim}{Claim}[section] 

\usepackage{balance}

\usepackage{epsfig}

\usepackage{graphicx}

\usepackage{amsmath}
\usepackage{amsfonts}

\usepackage{tabularx}
\usepackage{booktabs}
\usepackage{relsize}

\usepackage{numprint} 

\usepackage{setspace}
\graphicspath{{./}{./graphics/}}
\newcolumntype{H}{>{\setbox0=\hbox\bgroup}c<{\egroup}@{}}

\newcommand{\abs}[1]{\left|#1\right|}

\usepackage{algorithm}
\usepackage{algorithmicx}
\usepackage{algpseudocode}
\algblockdefx[parallel]{ParFor}{EndPar}[1][]{$\textbf{parallel for}$ #1 $\textbf{do}$}{$\textbf{end parallel}$}
\algrenewcommand{\alglinenumber}[1]{\fontsize{6.5}{7}\selectfont#1}
\algtext*{EndPar}

\algblockdefx[parallel]{parfor}{endpar}[1][]{$\textbf{parallel for}$ #1 $\textbf{do}$}{$\textbf{end parallel}$}
\algrenewcommand{\alglinenumber}[1]{\scriptsize#1:}

\algblockdefx[parallel]{parallelfor}{parallelend}
[1][]{\textbf{parallel for} #1}
{\textbf{end parallel}}

\usepackage{nicefrac}

\algnotext{EndFor}
\algnotext{EndIf}
\algnotext{EndProcedure}

\DeclareMathAlphabet{\mathbcal}{OMS}{cmsy}{b}{n}

\definecolor{dkgreen}{rgb}{0,0.6,0}
\definecolor{gray}{rgb}{0.5,0.5,0.5}
\definecolor{lightred}{rgb}{0.93,0.93,0.93}
\definecolor{lightred}{rgb}{0.83,0.83,0.83} 

\definecolor{lightgraydark}{rgb}{0.6,0.6,0.6}

\definecolor{lightblue}{rgb}{0.5,0.90,1.0}
\definecolor{lightgreen}{rgb}{0.5,0.92,0.5}
\definecolor{lightyellow}{rgb}{1,0.90,0.40}

\definecolor{verylightgreen}{RGB}	{204,255,204}
\definecolor{verylightred}{RGB}		{255,204,204}
\definecolor{verylightyellow}{RGB}		{255,255,204}

\definecolor{lightgreen}{RGB}	{204,255,204}
\definecolor{lightyellow}{RGB}		{255,255,204}

\definecolor{plotblue}{RGB}	{30,144,255}
\definecolor{plotgreen}{RGB}	{50,205,50}
\definecolor{plotred}{RGB}	{220,20,60}

\definecolor{myyellow}{RGB}{255,255,204}
\definecolor{myred}{RGB}{255,204,204}
\definecolor{lightblue}{RGB}{0,200,255}
\definecolor{mygreen}{RGB}{80,220,80}

\definecolor{gray}{RGB}{20,20,20}
\definecolor{greencm}{RGB}{0,153,0}

\definecolor{theblue}{RGB}{0,0,180}

\definecolor{matlabgreen}{rgb}{0,0.6,0}
\definecolor{matlabgray}{rgb}{0.5,0.5,0.5}
\definecolor{matlabmauve}{rgb}{0.58,0,0.82}

\usepackage[notheorems]{rr-math}
\renewcommand{\vv}{\ensuremath{\vec{v}}}

\begin{document}
\title{Personalized Visualization Recommendation}

\author{Xin Qian}
\orcid{1234-5678-9012-3456}
\affiliation{
\institution{University of Maryland}
\city{College Park}
\state{MD}
\country{USA}
}
\email{xinq@umd.edu}

\author{Ryan A. Rossi}
\orcid{1234-5678-9012-3456}
\affiliation{
\institution{Adobe Research}
\city{San Jose}
\state{CA}
\country{USA}
}
\email{rrossi@adobe.com}

\author{Fan Du}
\affiliation{
\institution{Adobe Research}
\city{San Jose}
\state{CA}
\country{USA}
}
\email{fdu@adobe.com}

\author{Sungchul Kim}
\affiliation{
\institution{Adobe Research}
\city{San Jose}
\state{CA}
\country{USA}
}
\email{skim@adobe.com}

\author{Eunyee Koh}
\affiliation{
\institution{Adobe Research}
\city{San Jose}
\state{CA}
\country{USA}
}
\email{eunyee@adobe.com}

\author{Sana Malik}
\affiliation{
\institution{Adobe Research}
\city{San Jose}
\state{CA}
\country{USA}
}
\email{sana.malik@adobe.com}

\author{Tak Yeon Lee}
\affiliation{
\institution{Adobe Research}
\city{San Jose}
\state{CA}
\country{USA}
}
\email{talee@adobe.com}

\author{Nesreen K. Ahmed}
\affiliation{
\institution{Intel Labs}
\city{Santa Clara}
\state{CA}
\country{USA}
}
\email{nesreen.k.ahmed@intel.com}

\renewcommand{\shortauthors}{X. Qian et al.}

\begin{abstract}
Visualization recommendation work has focused solely on scoring visualizations based on the underlying dataset, and not the actual \emph{user} and their past visualization feedback. These systems recommend the same visualizations for every user, despite that the underlying user interests, intent, and visualization preferences are likely to be fundamentally different, yet vitally important. In this work, we formally introduce the problem of \emph{personalized visualization recommendation} and present a generic learning framework for solving it. In particular, we focus on recommending visualizations personalized for each individual user based on their past visualization interactions (\eg, viewed, clicked, manually created) along with the data from those visualizations. More importantly, the framework can learn from visualizations relevant to other users, even if the visualizations are generated from completely different datasets. Experiments demonstrate the effectiveness of the approach as it leads to higher quality visualization recommendations tailored to the specific user intent and preferences. To support research on this new problem, we release our user-centric visualization corpus consisting of 17.4k users exploring 94k datasets with 2.3 million attributes and 32k user-generated visualizations.
\end{abstract}

\maketitle

\section{Introduction}
With massive datasets becoming ubiquitous, visualization recommendation systems have become increasingly important.
These systems have the promise of enabling rapid visual analysis and exploration of such datasets.
However, existing end-to-end visualization recommendation systems output a long list of visualizations based solely on simple visual rules~\cite{voyager,voyager2}.
These systems lack the ability to recommend visualizations that are personalized to the specific user and the tasks that are important to them.
This makes it both time-consuming and difficult for users to effectively explore such datasets and find meaningful visualizations. 

Recommending visualizations that are \emph{personalized} to a specific user is an important unsolved problem.
Prior work on visualization recommendation have focused mainly on rule-based or ML-based approaches that are completely agnostic to the user of the system. In particular, these systems recommend the same ranked list of visualizations for every user, despite that the underlying user interests, intent, and visualization preferences are fundamentally different, yet vitally important for recommending useful and interesting visualizations for a specific user.
The rule-based methods use simple visual rules to score visualizations whereas the existing ML-based methods have focused solely on classifying design choices~\cite{vizml} or ranking such design choices~\cite{draco} using a corpus of visualizations that are \emph{not} tied to a user. 
Neither of these existing classes of visualization recommendation systems focus on modeling individual user behavior nor personalizing for individual users, which is at the heart of our work.

In this work, we introduce a new problem of \emph{personalized visualization recommendation} and propose an expressive framework for solving it.
The problem studied in this work is as follows:
Given a set of $n$ users where each user has their own specific set of datasets, and each of the user datasets contains a set of relevant visualizations (\ie, visualizations a specific user has interacted with in the past, 
either implicitly by clicking/viewing or 
explicitly by liking or adding the visualization to their favorites or a dashboard they are creating),
the problem of \emph{personalized visualization recommendation} is to learn an individual recommendation model for every user such that when a user selects a possibly new dataset of interest, 
we can apply the model for that specific user to recommend the top most relevant visualizations that are most likely to be of interest to them 
(despite that there is no previous implicit/explicit feedback on any of the visualizations from the new dataset).
Visualizations are fundamentally tied to a dataset as they consist of the 
(i) set of visual design choices (\eg, chart-type, color/size, x/y) and the
(ii) subset of data attributes from the full dataset used in the visualization.
Therefore, how can we develop a learning framework for solving the personalized visualization recommendation problem that is able to learn from other users and their relevant visualizations, even when those visualizations 
are from tens of thousands of completely different datasets with no shared attributes?

There are two important and fundamental issues at the heart of the personalized visualization recommendation problem.
First, since visualizations are defined based on the attributes within a single specific dataset, there is no way to leverage the visualization preferences of users across different datasets.
Second, since each user often has their own dataset of interest (not shared by other users),
there is no way to leverage user preferences across different datasets.
In this work, we address both problems.
Notably, the framework proposed in this paper
naturally generalizes to the following problem settings:
(a) single dataset with a single set of visualizations shared among all users, and
(b) tens of thousands of datasets 
that are not shared between users where each dataset of interest to a user gives rise to a completely different set of possible visualizations.
However, the existing work cannot be used to solve the new 
problem formulation that relaxes the single dataset assumption to make it more general and widely applicable.

In the problem formulation of personalized visualization recommendation, each user can have their own
set of datasets, and since each visualization represents a series of design choices and data (\ie, attributes tied to a specific dataset), then this gives rise to a completely disjoint set of visualizations for each user.
Hence, there is no way to directly leverage visualization feedback from other users, since the visualizations are from different datasets.
Furthermore, visualizations are dataset specific, since they are generated based on the underlying dataset, and therefore any feedback from a user cannot be directly leveraged for making better recommendations for other users and datasets.
To understand the difficulty of the proposed problem of personalized visualization recommendation, the equivalent problem with regards to traditional recommender systems would be as if each user on Amazon (or Netflix) had their own separate set of disjoint products (or movies) that no other user could see and provide feedback.
In such a setting, 
how can we then use feedback from other users?
Furthermore, given a single dataset uploaded by some user, there are an exponential number of possible visualizations that can be generated from it.
This implies that even if there are some users interested in a single dataset, the amount of preferences by those users is likely to be extremely small compared to the exponential number of possible visualizations that can be generated and preferred by such users.

To overcome these issues and make it possible to solve the personalized visualization recommendation problem,
we introduce two new models and representations that enable learning from dataset and visualization preferences across different datasets and users, respectively.
First, we propose a novel model and representation that encodes users, their interactions with attributes (i.e., attributes in any dataset) and we map every attribute to a shared k-dimensional meta-feature space that enables the model to learn from \emph{user-level data preferences} across all the different datasets of the users.
Most importantly, the shared meta-feature space is independent of the specific datasets and the meta-features represent general functions of an arbitrary attribute, independent of the user or dataset that it arises.
This enables the model to learn from user-level data preferences, despite that those preferences are on entirely different datasets.
Second, we propose a novel \emph{user-level visual preference} graph model for visualization recommendation using the proposed notion of a visualization configuration that enables learning from user-level visual preferences across different datasets and users.
Importantly, the graph model is able to directly learn from user-level visual preferences across different datasets.
This model encodes users and their visual-configurations (sets of design choices). 
Since each visual-configuration node represents a set of design choices that are by definition not tied to a user-specific dataset, then the proposed model can use this user-level visual graph to infer and make connections between other similar visual-configurations that are also likely to be useful to that user. 
This new graph model is critical since it allows the learning component to learn from user-level visual preferences (which are visual-configurations) across the different datasets and users.
Without this novel component, there would be no way to learn from other users visual preferences (sets of design choices).

\subsection{Summary of Contributions}
This work makes the following key contributions:
\begin{itemize}
\item \textbf{Problem Formulation:} 
We introduce and formulate the problem of \emph{personalized visualization recommendation} that learns a personalized 
visualization recommendation model
for every individual user 
based on their past visualization feedback, and the feedback of other users and their relevant visualizations from completely different datasets.
Our formulation removes the unrealistic assumption of a single dataset shared across all users (and thus that there exists a single set of dataset-specific visualizations shared among all users).
To solve this problem, the model must be able to learn from the visualization and data preferences of many users across tens of thousands of different datasets.

\item \textbf{Framework:} 
We propose a flexible framework that expresses a class of methods for the
\emph{personalized visualization recommendation} problem. 
To solve this new problem, we introduce new graph representations and models that enable learning from the visualization and data preferences of users despite them being in different datasets entirely.
More importantly, the proposed framework is able to exploit the visualization and data preferences of users across tens of thousands of different datasets.

\item \textbf{Effectiveness:} 
The extensive experiments demonstrate the importance and effectiveness of learning personalized visualization recommendation models for each individual user.
Notably, our personalized models perform significantly better than SOTA baselines with a mean improvement of 29.8\% and 64.9\% for HIT@5 and NDCG@5, respectively.
Furthermore, the deep personalized visualization recommendation models are shown to perform even better.
Finally, comprehensive ablation studies are performed to understand the effectiveness of the different learning components.

\end{itemize}

\subsection{Organization of article}
First, we introduce a new problem of visualization recommendation in Section~\ref{sec:problem} that learns a personalized model for each of the $n$ individual users by leveraging a large collection of datasets and relevant visualizations from each of the datasets in the collection.
Notably, the learning of the individual user models are able to exploit the preferences of other users (even if the preferences are on a completely different dataset) including the data attributes used in a visualization, visual design choices, and actual visualizations generated despite that no other user may have used the underlying dataset of interest.
In Section~\ref{sec:framework}, we propose a computational framework for solving the new problem of personalized visualization recommendation.
Further, we also propose \emph{deep personalized visualization recommendation models} in Section~\ref{sec:neural-pvisrec} that are able to learn complex non-linear functions between the embeddings of the users, visualization-configurations, datasets and the data attributes used in the visualizations.
Next, Section~\ref{sec:dataset} describes the user-centric visualization corpus we created and made publicly accessible
for studying this problem.
Then Section~\ref{sec:exp} provides a comprehensive and systematic evaluation of the proposed approach and framework for the personalized visualization recommendation problem 
while Section~\ref{sec:related-work} discusses related work.
Finally, Section~\ref{sec:conc} concludes with a summary of the key findings and briefly discusses directions for future work on this new problem.

\section{Personalized Visualization Recommendation} \label{sec:problem}
In this section, we formally introduce the \emph{Personalized Visualization Recommendation} problem.
The personalized visualization recommendation problem has two main parts: 
(1) training a personalized visualization recommendation model for every user $i \in [n]$ (Section~\ref{sec:problem-model-training}), and
(2) leveraging the user-personalized model to recommend personalized visualizations based on the users past dataset and visualization feedback/preferences (Section~\ref{sec:problem-recommending-personalized-vis}).
\begin{enumerate}
\item \textbf{Personalized Model Training (Sec.~\ref{sec:problem-model-training}):} 
Given a user-level training visualization corpus $\mathbcal{D} = \{(\mathcal{X}_{i}, \mathbb{V}_{i})\}_{i=1}^n$ consisting of $n$ users and their corresponding datasets of interest $\mathbcal{X}_{i}=\{\mX_{i1},\ldots,\mX_{ij},\ldots\}$ as well as their relevant sets of visualizations $\mathbb{V}_{i}=\{\mathbcal{V}_{i1},\ldots,\mathbcal{V}_{ij},\ldots\}$ for those datasets, 
we first learn a user-level personalized model $\mathcal{M}$ from the training corpus $\mathbcal{D}$ that best captures and scores the effective visualizations for user $i$ highly while assigning low scores to visualizations that are likely to not be preferred by the user.

\item \textbf{Recommending Personalized Visualizations (Sec.~\ref{sec:problem-recommending-personalized-vis}):} 
Given a user $i \in [n]$ and a dataset $\mX_{ij}$ of interest to user $i$, 
we use the trained personalized visualization recommendation model $\mathbcal{M}$ for user $i$ 
to generate, score, and recommend the top visualizations of interest to user $i$ for dataset $\mX_{ij}$.
Note that we naturally support the case when the dataset $\mX_{ij} \not\in \mathbcal{X}_i$ is new or when the dataset $\mX_{ij} \in \mathbcal{X}_i$ is not new, but we have at least one or more previous user feedback about the visualizations the user likely prefers from that dataset.
\end{enumerate}
The fundamental difference between the ML-based visualization recommendation problem introduced in~\cite{ML-based-Vis-Rec} and the personalized visualization recommendation problem described above is that the personalized problem focuses on modeling the behavior, data, and visualization preferences of individual users.
Since local visualization recommendation models are learned for every user $i \in [n]$ (as opposed to training a global visualization recommendation model),
it becomes important to leverage every single piece of feedback from the users.
For instance, global visualization recommendation models essentially ignore the notion of a user, and therefore can leverage all available training data to learn the best global visualization recommendation model.
However, personalized visualization recommendation models explicitly leverage specific user feedback to learn the best personalized local model for every user $i \in [n]$, and there of course is far less feedback from individual users.

\subsection{Implicit and Explicit User Feedback for Personalized Vis. Rec.}
In this work, relevant visualizations $\mathbcal{V}_{ij} \in \mathbb{V}_i$ for a specific user $i$ and dataset $\mX_{ij} \in \mathbcal{X}_i$ are defined generally, as the term relevant may refer to visualizations that a user clicked, liked, generated, among many other user actions that demonstrate positive feedback towards a visualization.
In terms of personalized visualization recommendation, there are two general types of user feedback: implicit or explicit user feedback.
Implicit user visualization feedback corresponds to user feedback that is not explicitly stated and includes user actions such as when a user clicks on a visualization or hovers over a visualizations for more than a specific time.
Conversely, explicit user feedback on a visualizations refers to feedback that is more explicitly stated about a visualization such as when a user explicitly likes a visualizations, or generates a visualization.
Obviously, implicit user feedback is available at a larger quantity than explicit user feedback.
However, implicit user feedback is not as strong as user feedback that is explicit, \eg, a user that clicked a visualization is not as strong as a user that explicitly liked a visualization.

We propose two different types of user preferences (implicit and explicit user feedback) that are important for learning personalized visualization recommendation models for individual users, including
the data preferences \emph{and} visual preferences of each individual user.
For learning the data and visual preferences of a user, there is both implicit and explicit user feedback that can be used for developing personalized visualization recommender systems.

\subsubsection{Data Preferences of Users: Implicit and Explicit Data Feedback}
There is naturally both implicit \emph{and} explicit user feedback regarding the \emph{data preferences} of users.
Explicit user feedback about the data preferences of a user is a far stronger signal than implicit user feedback, however, there is typically a lot more implicit user feedback for learning than explicit feedback from the user.
\begin{itemize}
\item \textbf{Implicit Data Preferences of Users.}
An example of \emph{implicit feedback w.r.t. data preferences of the user} is when a user clicks (or hovers over) a visualization that uses two attributes $\vx$ and $\vy$ from some arbitrary user-selected dataset.
We can then extract the users data preferences from the visualization by encoding the two attributes that were used in the visualization preferred by that user.

\item \textbf{Explicit Data Preferences of Users.}
Similarly, an example of \emph{explicit feedback w.r.t. data preferences of the user} is when a user explicitly likes a visualization (or adds a visualization to their dashboard) that uses two attributes $\vx$ and $\vy$ from some arbitrary user-selected dataset.
In this work, we use another form of explicit feedback based on a user-generated visualization and the attributes (data) used in the generated visualization.
Hence, this is a form of explicit feedback, since the user explicitly selects the attributes and creates a visualization using them (as opposed to clicking on a visualization automatically generated by a system).
\end{itemize}\noindent
Besides using implicit and explicit feedback provided by the user based on the click or like of a visualization and the data used in it, we can also leverage an even more direct feedback about a users data preferences.
For instance, many visualization recommender systems allow users to select an attribute of interest to use in the recommended visualizations. 
As such we can naturally leverage any feedback of this type as well.

\subsubsection{Visual Preferences of Users: Implicit and Explicit Visual Feedback}
In terms of visual preferences of users, there is both implicit \emph{and} explicit user feedback that can be used to learn a better personalized visualization recommendation model for individual users.
\begin{itemize}
\item \textbf{Implicit Visual Preferences of Users.}
An example of \emph{implicit feedback w.r.t. visual preferences of the user} is when a user clicks (or hovers over) a visualization from some arbitrary user-selected dataset.
We can then extract the users visual preferences from the visualization, and appropriately encode it for learning the visual preferences of the individual user.

\item \textbf{Explicit Visual Preferences of Users.}
Similarly, an example of \emph{explicit feedback w.r.t. visual preferences of the user} is when a user explicitly likes a visualization (or adds a visualization to their dashboard).
Just as before, we can then extract the visual preferences of the user from the visualization (mark/chart type, x-type, y-type, color, size, x-aggregate, and so on) and leverage the individual visual preferences or a combination of them for learning the user-specific personalized vis. rec. model.
\end{itemize}

\subsection{Training User Personalized Visualization Recommendation Model} \label{sec:problem-model-training}
Given user log data 
\begin{align} \label{eq:user-feedback-log-data}
\mathbcal{D} &= \{(\mathbcal{X}_1, \mathbb{V}_1),\ldots,(\mathbcal{X}_i, \mathbb{V}_i),\ldots,(\mathbcal{X}_n, \mathbb{V}_n)\} = \{(\mathbcal{X}_i, \mathbb{V}_i)\}_{i=1}^n
\end{align}
where for each user $i \in [n]$, we have the set of datasets of interest to that user denoted as $\mathbcal{X}_i$ along with the sets of relevant visualizations $\mathbb{V}_i$ generated by user $i$ for every dataset $\mX_{ij} \in \mathbcal{X}_i$.
More specifically, 
\begin{align}
\mathbb{V}_{i} &= \{\mathbcal{V}_{i1}, \ldots, \mathbcal{V}_{ij}, \ldots \} 
\quad\text{ and }\;\;\;\;
\mathbcal{V}_{ij} = \{\mathbcal{V}_{ij1},\ldots,\mathcal{V}_{ijk},\ldots\}
\\
\mathbcal{X}_{i} &= \{\mX_{i1}, \ldots, \mX_{ij}, \ldots \} 
\quad\text{ and }\;\;\;\;
\mX_{ij} = [\vx_{ij1}\,\, \vx_{ij2}\, \cdots]
\end{align}
where $\vx_{ijk}$ is the $k$th attribute (column vector) of $\mX_{ij}$.
Hence, the number of attributes in $\mX_{ij}$ has no relation to the number of relevant visualizations $|\mathbcal{V}_{ij}|$ that a user $i$ preferred for that dataset.
For a single user $i$, the number of user preferred visualizations across all datasets of interest for that user is
\begin{equation}
v_i = \sum_{\mathbcal{V}_{ij} \in \mathbb{V}_i} |\mathbcal{V}_{ij}|
\end{equation}
where $\mathbcal{V}_{ij}$ is the set of visualizations preferred by user $i$ from dataset $j$.
Thus, the total number of user generated visualizations across all users and datasets is
\begin{equation}
v = \sum_{i=1}^n \sum_{\mathbcal{V}_{ij} \in \mathbb{V}_i} |\mathbcal{V}_{ij}|
\end{equation}
For simplicity, let $\mathcal{V}_{ijk} \in \mathbcal{V}_{ij} = \{\mathcal{V}_{ij1},\ldots,\mathcal{V}_{ijk},\ldots\}$ denote the visualization generated by user $i$ from dataset $j$, that is, $\mX_{ij} \in \mathbcal{X}_i$, specifically using the subset of attributes $\mX_{ij}^{(k)}$ from the dataset $\mX_{ij}$.
Further, every user $i \in [n]$ is associated with a set of datasets $\mathbcal{X}_i = \{\mX_{i1},\ldots,\mX_{ij}, \ldots\}$ of interest.
Let $\mX_{ij}$ be the $j$th dataset of interest for user $i$ and let $|\mX_{ij}|$ denote the number of attributes (columns) of the dataset matrix $\mX_{ij}$.
Then the number of attributes across all datasets of interest to user $i$ is
\begin{align}
m_i = \sum_{\mX_{ij} \in \mathbcal{X}_i} |\mX_{ij}|
\end{align}
and the number of attributes across all $n$ users and all their datasets is
\begin{align}
m = \sum_{i=1}^{n} \sum_{\mX_{ij} \in \mathbcal{X}_i} |\mX_{ij}|
\end{align}

\begin{Definition}[Space of Attribute Combinations]\label{def:attribute-combination-space}
Given an arbitrary dataset matrix $\mX_{ij}$, let $\mathbb{X}_{ij}$ denote the space of attribute combinations of $\mX_{ij}$ defined as
\begin{align}
\Sigma : \mX_{ij} \to \mathbb{X}_{ij}, \quad \text{s.t.} \label{eq:attr-comb-generator-func} \\
\mathbb{X}_{ij} = \{\mX_{ij}^{(1)},\ldots,\mX_{ij}^{(k)},\ldots\}, \label{eq:attr-combinations-of-a-dataset}
\end{align}\noindent
where $\Sigma$ is an attribute combination generation function 
and every $\mX_{ij}^{(k)} \in \mathbb{X}_{ij}$ is a different subset (combination) of attributes from $\mX_{ij}$ consisting of one or more attributes from $\mX_{ij}$.
\end{Definition}\noindent
\begin{Property}\label{prop:comparing-attr-comb-spaces}
Let $|\mX_{ij}\!|$ and $|\mX_{ik}|$ denote the number of attributes (columns) of two arbitrary datasets $|\mX_{ij}|$ and $|\mX_{ik}|$ of user $i$.
If $|\mX_{ij}|>|\mX_{ik}|$, then $|\mathbb{X}_{ij}|>|\mathbb{X}_{ik}|$.
\end{Property}\noindent
It is straightforward to see that if $|\mX_{ij}| > |\mX_{ik}|$, then the number of attribute combinations of $\mX_{ij}$ denoted as $|\mathbb{X}_{ij}|$ is larger than the number of different attribute subsets that can be generated from $\mX_{ik}$ denoted as $|\mathbb{X}_{ik}|$.
Property~\ref{prop:comparing-attr-comb-spaces} is important as it characterizes the space of attribute combinations/subsets for a given dataset $\mX_{ij}$ and therefore can be used to understand the corresponding space of possible visualizations that can be generated from a given dataset, as these are also tied.

In this work, we assume a visualization is specified using some grammar such as Vega-Lite~\cite{vega-lite}.
Therefore, the data mapping and design choices of the visualization are encoded in json (or json-like format), and can easily render a visualization.
A visualization configuration $\mathcal{C}$ (design choices) \emph{and} the data attributes $\mX^{(k)}_{ij}$ selected from a dataset $\mX_{ij}$ is everything necessary to generate a visualization $\mathcal{V}=(\mX^{(k)}_{ij},\mathcal{C})$.
Hence, the tuple $(\mX^{(k)}_{ij},\mathcal{C})$ defines a unique visualization $\mathcal{V}$ that leverages the subset of attributes $\mX^{(k)}_{ij}$ from dataset $\mX_{ij}$ along with the visualization configuration $\mathcal{C} \in \mathbcal{C}$.
\begin{Definition}[Visualization Configuration] \label{def:vis-config}
Given a visualization $\mathcal{X}$ generated using a subset of attributes $\mX_{ij}^{(k)}$ from dataset $\mX_{ij}$, 
we define a function
\begin{align}
\Gamma : \mathcal{V}\to \mathcal{C}
\end{align}\noindent
where $\Gamma$ maps every data-dependent 
design choice of the visualization to its corresponding type (\ie, the attribute mapping to the x-axis of the visualization $\mathcal{V}$ is replaced with its general type such as quantitative, nominal, ordinal, temporal, etc).
The resulting visualization configuration $\mathcal{C}$ is an abstraction of the visualization $\mathcal{V}$, in the sense that all the data attribute bindings have been abstracted and replaced with their general data attribute type.
Hence, $\mathcal{C}$ is an abstraction of $\mathcal{V}$.
\end{Definition}
\begin{Definition}[Space of Visualization Configurations] \label{def:space-vis-configs}
Let $\mathbcal{C}$ denote the space of all visualization configurations such that 
a visualization configuration $\mathcal{C}_{ik} \in \mathbcal{C}$ defines
an abstraction of a visualization where for each visual design choice (x, y, marker-type, color, size, etc.) that maps to an attribute in some  dataset $\mX_{ij}$, we replace it with its type such as quantitative, nominal, ordinal, temporal or some other general property characterizing the attribute that can be selected.
Therefore visualization configurations are essentially visualizations without any attributes (data), or visualization abstractions that are by definition data-independent.
\end{Definition}

\begin{Property}\label{prop:visualization-configuration-independent}
Every visualization configuration $\mathcal{C}_{ik} \in \mathbcal{C}$ is independent of any data matrix $\mX$ (by Definition~\ref{def:space-vis-configs}).
\end{Property}
The above implies that $\mathcal{C}_{ik} \in \mathbcal{C}$ can potentially arise from any arbitrary dataset and is therefore not tied to any specific dataset since visualization configurations are general abstractions where the data bindings have been replaced with their general type, \eg, if x/y in some visualization mapped to an attribute in $\mX$, then it is replaced by 
its type (\ie, ordinal, quantitative, categorical, etc).
A visualization configuration \emph{and} the attributes selected from a dataset is everything necessary to generate a visualization.
The size of the space of visualization configurations is large since visualization configurations come from all possible combinations of design choices and their values.

\begin{Definition}[Space of Visualizations of $\mX_{ij}$] \label{def:space-of-visualizations}
Given an arbitrary dataset matrix $\mX_{ij}$, we define $\mathbb{V}^{\star}_{ij}$ as the space of all possible visualizations that can be generated from $\mX_{ij}$.
More formally, the space of visualizations $\mathbb{V}^{\star}_{ij}$ is defined with respect to a dataset $\mX_{ij}$ and the space of visualization configurations $\mathbcal{C}$, 
\begin{align} 
&\mathbb{X}_{ij} = \Sigma(\mX_{ij}) = \{\mX_{ij}^{(1)},\ldots,\mX_{ij}^{(k)},\ldots\}\\
&\xi : \mathbb{X}_{ij} \times \mathbcal{C} \to \mathbcal{V}_{ij}^{\star} \label{eq:vis-generator}
\end{align}\noindent
where $\mathbb{X}_{ij} = \{\mX_{ij}^{1},\ldots,\mX_{ij}^{(k)},\ldots\}$ is the set of all possible attribute combinations of $\mX_{ij}$ (Def.~\ref{def:attribute-combination-space}).
More succinctly, $\xi : \Sigma(\mX_{ij}) \times \mathbcal{C} \to \mathbcal{V}_{ij}^{\star}$, and therefore  $\xi(\Sigma(\mX_{ij}),\mathbcal{C}) = \mathbcal{V}_{ij}^{\star}$.
The space of all visualizations $\mathbcal{V}_{ij}^{\star}$ is determined entirely by the underlying dataset, and therefore remains the same for all $n$ users.
The difference in our personalized visualization recommendation problem is the relevance of each visualization in the space of all possible visualizations generated from an arbitrary dataset.
Given a subset of attributes $\mX_{ij}^{(k)} \!\in \mX_{ij}$ from dataset $\mX_{ij}$ and a visualization configuration $\mathcal{C} \in \mathbcal{C}$, then 
$\xi(\mX_{ij}^{(k)}\!, \mathcal{C}) \in \mathbcal{V}_{ij}^{\star}$ is the corresponding visualization.
\end{Definition}\noindent
Importantly, fix $\mathbcal{C}$ and 
let $\mX \not= \mY \implies \forall i,j\,\, \vx_i \not= \vy_j$, then 
$\xi(\Sigma(\mX),\mathbcal{C}) \cap \xi(\Sigma(\mY),\mathbcal{C})=\emptyset$.
This implies the space of possible visualizations that can be generated is entirely dependent on the dataset (not the user).
Hence, for any two datasets $\mX$ and $\mY$ without any shared attributes between them,
the set of visualizations that can be generated from $\mX$ or $\mY$ is completely different,
\[\xi(\Sigma(\mX),\mathbcal{C}) \cap \xi(\Sigma(\mY),\mathbcal{C})=\emptyset\]
This has important consequences for the new problem of personalized visualization recommendation.
Since it is unlikely that any two users care about the same underlying dataset, and even if they did, it is even far more unlikely that they have any relevant visualizations in common 
(just \wrt the exponential size of the visualization space for a single dataset with a reasonable amount of attributes).
Therefore, it is not possible nor practical to leverage the relevant visualizations of a user directly.
Instead, we need to decompose a visualization $\mathcal{V}$ into its more meaningful components such as: 
(i) the \emph{characteristics} of the data attributes $\mX_{ij}^{(k)}$ used in a visualization, and 
(ii) the visual design choices (chart-type/mark, color, size, and so on).

\begin{Definition}[Relevant Visualizations of User $i$ and Dataset $\mX_{ij}$]
Let $\mathbcal{V}_{ij} \in \mathbb{V}_i$ define the set of relevant (positive) visualizations for user $i$ with respect to dataset $\mX_{ij}$.
Therefore, 
$\mathbb{V}_i = \bigcup_{\mX_{ij} \in \mathbcal{X}_i} \mathbcal{V}_{ij}$
where $\mathbb{V}_i$ is the set of all relevant visualizations across all datasets $\mathbcal{X}_i$ of interest to user $i$.
\end{Definition}

\begin{Definition}[Non-relevant Visualizations of User $i$ and Dataset $\mX_{ij}$] \label{def:negative-vis-of-a-dataset}
For a user $i$, let $\mathbb{V}_{ij}^{\star}$ denote the space of all visualizations that arise from the $j$th dataset $\mX_{ij}$ such that the relevant (positive) visualizations $\mathbcal{V}_{ij}$ satisfies $\mathbcal{V}_{ij} \subseteq \mathbcal{V}_{ij}^{\star}$, then the \emph{space of non-relevant visualizations for user $i$ on dataset $\mX_{ij}$} is 
$\mathbcal{V}_{ij}^{-} = \mathbcal{V}_{ij}^{\star} \setminus \mathbcal{V}_{ij}$, which follows from $\mathbcal{V}_{ij}^{-} \cup \mathbcal{V}_{ij} = \mathbcal{V}_{ij}^{\star}$.
\end{Definition}

\begin{table}[b!]
\caption{Summary of notation. Matrices are bold upright roman letters; vectors are bold lowercase letters.}
\vspace{-3mm}
\centering 
\fontsize{8}{8.5}\selectfont
\setlength{\tabcolsep}{6pt} 
\label{table:notation}
\def\arraystretch{1.3}
\begin{tabularx}{1.00\linewidth}{
c X} 
\toprule$\mathbcal{D}$ & user log data $\mathbcal{D}=\{(\mathbcal{X}_i,\mathbb{V}_i)\}_{i=1}^{n}$ consisting of a set of datasets $\mathbcal{X}_i$ for every user $i \in [n]$ and the sets $\mathbb{V}_i$ of relevant visualizations for each of those datasets.\\
$\mathbcal{X}_i$ & set of datasets (data matrices) of interest to user $i$ where $\mathbcal{X}_i = \{\mX_{i1}, \ldots, \mX_{ij},\ldots\}$\\
$\mX_{ij}$ & the $j$th dataset (data matrix) of interest to user $i$.\\
$\mathbb{V}_i$ & sets of visualizations relevant to user $i$ where $\mathbb{V}_i = \{\mathbcal{V}_{i1}, \ldots, \mathbcal{V}_{ij}, \ldots\}$\\
$\mathbcal{V}_{ij}$ & set of visualizations relevant (generated) by user $i$ for dataset $j$ ($\mX_{ij} \in \mathbcal{X}_i$) where $\mathbcal{V}_{ij}=\{\ldots,\mathcal{V},\}$\\
$\mathcal{V}=(\mX^{(k)}, C)$ & a visualization $\mathcal{V}$ consisting of the subset of attributes $\mX^{(k)}$ from some dataset $\mX$ and the visual-configuration (design choices) \\
$\mathbcal{C}$ & set of visual-configurations where $\mathcal{C} \in \mathbcal{C}$ represents the visualization design choices for a single visualization $\mathcal{V}$ such as the chart-type, x-axis, y-axis, color, and so on. \\
$\mathbb{X}_{ij}$ & space of attribute combinations/subsets $\mathbb{X}_{ij} = \{\mX_{ij}^{(1)},\ldots,\mX_{ij}^{(k)},\ldots\}$ of dataset $\mX_{ij}$ \\
\midrule
$n$ & number of users \\
$m$ & number of attributes (columns, variables) across all datasets, $m = \sum_{i} m_i$ 
where $m_i$ = number of attributes in the $i$-th dataset \\
$v$ & number of relevant (user-generated) visualizations across all users and datasets \\
$h$ & number of visualization configurations \\
$k$ & dimensionality of the \emph{shared attribute feature space}, \ie, number of attribute features \\
$d$ & shared latent embedding dimensionality \\
$t$ & number of types of implicit/explicit user feedback, \ie, attribute and visualization click, like, add-to-dashboard, among others \\ 
\midrule

$\vx$ & a attribute (column) vector from an arbitrary user uploaded dataset \\
$\abs{\vx}$ & cardinality of $\vx$, \ie, number of unique values in $\vx$ \\
$\mathsf{nnz}(\vx)$ & number of nonzeros in a vector $\vx$ \\
$\mathsf{len}({\vx})$ & length of a vector $\vx$ \\

\midrule
$\mA$  & user by attribute preference matrix \\ 
$\mC$  & user by visualization configuration matrix \\
$\mD$  & attribute preference by visual-configuration matrix \\
$\mM$ & attribute by meta-feature matrix \\

\midrule
$\mU$  & shared user embedding matrix \\ 
$\mV$  & shared attribute embedding matrix \\ 
$\mZ$  & shared visualization configuration embedding matrix \\ 
$\mY$  & meta-feature embedding matrix for the attributes across all datasets \\
\bottomrule
\end{tabularx}
\end{table}

\noindent
We denote $Y_{ijk}$ as the ground-truth label of a visualization $\mathcal{V}_{ijk} \in \mathbcal{V}_{ij}^{\star}$ where $Y_{ijk}=1$ if $\mathcal{V}_{ijk} \in \mathbcal{V}_{ij}$ and $Y_{ijk}=0$ otherwise.
Now we formulate the problem of training a user-level personalized visualization recommendation model $\mathbcal{M}_i$ for user $i$ from a large user-centric visualization training corpus $\mathbcal{D}$.
\begin{Definition}[Training \emph{Personalized} Vis. Recommendation Model] \label{def:personalized-vis-rec-model-learning-training}
Given the set of training datasets and relevant visualizations 
$\mathbcal{D} = \{(\mathbcal{X}_i, \mathbb{V}_i)\}_{i=1}^n$,
the goal is to learn a \emph{personalized} visualization recommendation model $\mathbcal{M}_i$ for user $i$ by solving 
the following general objective function,
\begin{align}\label{eq:vis-rec-user-model-learning}
\argmin_{\mathbcal{M}_i} 
\sum_{j=1}^{|\mathbcal{X}_i|} 
\sum_{(\mX_{ij}^{(k)}\!,\mathcal{C}_{ijk})\in \mathbcal{V}_{ij} \cup \hat{\mathbcal{V}}_{ij}^{-}}
\mathbb{L}\Big(
Y_{ijk} \,\big|\, 
\Psi(\mX_{ij}^{(k)}), 
f(\mathcal{C}_{ijk}), 
\mathbcal{M}_i
\Big), \quad i=1,\ldots,n
\end{align}\noindent
where $\mathbb{L}$ is the loss function,
$Y_{ijk}=\{0,1\}$ is the ground-truth label of the $k$th visualization $\mathcal{V}_{ijk} = (\mX_{ij}^{(k)}\!,\mathcal{C}_{ijk})\in \mathbcal{V}_{ij} \cup \hat{\mathbcal{V}}_{ij}^{-}$
for dataset $\mX_{ij} \in \mathbcal{X}_i$ of user $i$.
Further, $\mX_{ij}^{(k)} \subseteq \mX_{ij}$ is the subset of attributes used in the visualization.
In Eq.~\ref{eq:vis-rec-user-model-learning}, $\Psi$ and $f$ are general functions over the subset of attributes $\mX_{ij}^{(k)} \subseteq \mX_{ij}$ and the visualization configuration $\mathcal{C}_{ijk}$ of the visualization $\mathcal{V}_{ijk} = (\mX_{ij}^{(k)}\!,\mathcal{C}_{ijk}) \in \mathbcal{V}_{ij}^{-} \cup \mathbcal{V}_{ij}$, respectively.
\end{Definition}\noindent

For learning individual models $\mathbcal{M}_i$ for every user $i \in [n]$, we can also leverage the visualization and data preferences from other users.
The simplest and most straightforward situation is when there is another user $i^{\prime} \in [n]$ with a set of relevant visualizations that use attributes from the same exact dataset, hence $|\mathbcal{X}_i \cap \mathbcal{X}_{i^{\prime}}|>0$.
While the above strict assumption is convenient as it makes the problem far simpler, it is unrealistic in practice (and not very useful) to assume there exists a single dataset of interest to all users.
Therefore, we designed the approach to be able to learn from visualizations preferred by other users on completely different datasets.
This is done by leveraging the similarity between the attributes (used in the visualizations) across completely different datasets (by first embedding the attributes from every dataset to a shared fixed dimensional space) 
as well as the similarity between the visual-configurations of the relevant visualizations, despite them using completely different datasets.
More formally, given any two users $i, i^{\prime} \in [n]$ along with one of their relevant visualizations, 
$\mathcal{V}_{ijk} = (\mX_{ij}^{(k)}\!,\mathcal{C}_{ijk}) \in \mathbcal{V}_{ij}$
and $\mathcal{V}_{i^{\prime}\!j^{\prime}\!k^{\prime}} = (\mX_{i^{\prime}\!j^{\prime}}^{(k^{\prime})}\!,\mathcal{C}_{i^{\prime}\!j^{\prime}\!k^{\prime}}) \in \mathbcal{V}_{i^{\prime}\!j^{\prime}}$, 
then since we know that the datasets used in these visualizations are completely different, we instead can leverage this across-dataset training information if they use similar attributes, where across-dataset similarity is measured by first mapping each attribute used in the visualization to a shared $K$-dimensional meta-feature space, where we can then measure the similarity between each of the attributes used in the visualizations generated by different users.
Hence, $s\langle \Psi(\mX_{ij}^{(k)}), \Psi(\mX_{i^{\prime}\!j^{\prime}}^{(k^{\prime})}) \rangle > 1-\epsilon$
where $\mX_{i^{\prime}\!j^{\prime}} \not \in \mathbcal{X}_i$ and $\mX_{ij} \not \in \mathbcal{X}_{i^{\prime}}$.
Intuitively, this implies that even though the visualizations are generated using different data, they visualize data that is similar 
with respect to its overall characteristics and patterns.
By construction, visualizations $\mathcal{V}_{ijk}$ and $\mathcal{V}_{i^{\prime}\!j^{\prime}\!k^{\prime}}$ from two different users $i, i^{\prime} \in [n]$ and datasets $\mX_{ij} \not= \mX_{i^{\prime}\!j^{\prime}}$ may use the same visual-configuration (set of design choices), $\mathcal{C}_{ijk} = \mathcal{C}_{i^{\prime}\!j^{\prime}\!k^{\prime}} \in \mathbcal{C}$, since we defined the notion of visual-configurations to be data-independent,
and thus, even though two visualizations may visualize data attributes from completely different datasets, they can still share the same visual-configuration (design choices).
Therefore, as we will see later, we are able to learn from other users with visualizations that use attributes from completely different datasets.

\subsection{Personalized Visualization Scoring and Recommendation} \label{sec:problem-recommending-personalized-vis}

After learning the personalized visualization recommendation model $\mathbcal{M}_i$ for an individual user $i \in [n]$ (Eq.~\ref{eq:vis-rec-user-model-learning}), we can then use $\mathbcal{M}_i$ to score and recommend the top most relevant visualizations for user $i$ from any arbitrary dataset $\mX$.
There are three possible cases that are naturally supported by the learned model $\mathbcal{M}_i$ for recommending visualizations specifically of interest to user $i$ based on their past interactions (visualizations the user viewed/clicked or more generally interacted with):
\begin{compactenum}
\item The dataset $\mX$ used for recommending personalized visualizations to user $i$ via $\mathbcal{M}_i$ can be a new previously unseen dataset of interest 
$\mX \not\in \{\mathbcal{X}_1,\mathbcal{X}_2,\ldots,\mathbcal{X}_n\}$

\item The dataset $\mX$ is not a previous dataset of interest to user $i$, but has been used previously by one or more other users $\mX \in \{\mathbcal{X}_1,\ldots,\mathbcal{X}_n\} \setminus \mathbcal{X}_i$

\item The dataset $\mX \in \mathbcal{X}_i$ is a previous dataset of interest to user $i$
\end{compactenum}
A fundamental property of the personalized visualization recommendation problem is that the user visualization scores for an arbitrary visualization $\mathcal{V}$ (that visualizes data from an arbitrary dataset $\mX$) are different depending on the individual user and their historical preferences and interests.
More formally, 
given users $i, i^{\prime} \in [n]$ and a visualization $\mathcal{V}$ from a new unseen dataset $\mX_{\rm test}$, we obtain personalized visualization scores for user $i$ and $i^{\prime}$ as $\mathbcal{M}_i(\mathcal{V})$ and $\mathbcal{M}_{i^{\prime}}(\mathcal{V})$, respectively.
While existing rule-based~\cite{voyager,voyager2,draco} or ML-based systems~\cite{ML-based-Vis-Rec} score the visualization $\mathcal{V}$ the same, no matter the actual user of the system (hence, are agnostic to the actual user and their interests, past interactions, and intent),
our work instead focuses on learning individual personalized visualization recommendation models for every user $i\in [n]$ such that the personalized score $\mathbcal{M}_i(\mathcal{V})$ of visualization $\mathcal{V}$ for user $i$ is almost surely different from the score $\mathbcal{M}_{i^{\prime}}(\mathcal{V})$ given by the personalized model of another user $i^{\prime}$, $\mathbcal{M}_i(\mathcal{V}) \not= \mathbcal{M}_{i^{\prime}}(\mathcal{V})$.
We can state this more generally for all pairs of users $i, i^{\prime} \in [n]$ with respect to a single arbitrary visualization $\mathcal{V}$,
\begin{align}
\mathbcal{M}_i(\mathcal{V}) \not= \mathbcal{M}_{i^{\prime}}(\mathcal{V}),\quad \forall i, i^{\prime}=1,\ldots,n\quad \text{s.t.}\;\, i<i^{\prime}
\end{align}
\noindent
Hence, given an arbitrary visualization $\mathcal{V}$, the personalized scores $\mathbcal{M}_i(\mathcal{V})$ and $\mathbcal{M}_i^{\prime}(\mathcal{V})$ for any two distinct users $i$ and $i^{\prime}$ are not equal with high probability.
This is due to the fact that the personalized visualization recommendation models $\mathbcal{M}_1, \mathbcal{M}_2,\ldots,\mathbcal{M}_n$ capture each of the $n$ users individual data preferences, design/visual preferences, and overall visualization preferences.

\begin{Definition}[Personalized Visualization Scoring] \label{def:personalized-vis-scoring}
Given the personalized visualization recommendation model $\mathbcal{M}_i$ for user $i$ and a dataset $\mX_{\rm test}$ of interest to user $i$, we can obtain the personalized scores for user $i$ of every possible visualization that can be generated as,
\begin{equation}\label{eq:personalized-model-scoring}
\mathbcal{M}_i : \mathbcal{X}_{\rm test} \times \mathbcal{C} \to \RR
\end{equation}\noindent
where $\mathbcal{X}_{\rm test}=\{\ldots,\mX^{(k)}_{\rm test} ,\ldots\}$ is the space of attribute subsets from $\mX_{\rm test}$ and $\mathbcal{C}$ is the space of visualization configurations.
Hence, given an arbitrary visualization $\mathcal{V}$, the learned model $\mathbcal{M}_i$ outputs a personalized score for user $i$ describing the effectiveness or importance of the visualization with respect to that individual user.
\end{Definition}

\begin{Definition}[Personalized Visualization Ranking] \label{def:personalized-vis-ranking}
Given the set of generated visualizations 
$\mathbb{V}_{\rm test} = \{\mathbcal{V}_1, \mathbcal{V}_2, \ldots, \mathbcal{V}_{Q}\}$ where $Q = |\mathbb{V}_{\rm test}|$, 
we derive a personalized ranking of the visualizations $\mathbb{V}_{\rm test}$ from $\mX_{\rm test}$ for user $i$ as follows:
\begin{equation} \label{eq:personalized-ranking-vis}
\rho_i\big(\{\mathcal{V}_{1},\mathcal{V}_{2},\ldots,\mathcal{V}_{\mathcal{Q}}\}\big)\, =\; \argsort_{\mathcal{V}_{t} \in \mathbb{V}_{\rm test}} \; \mathbcal{M}_i(\mathcal{V}_{t})
\end{equation}\noindent
where for any two visualizations $\mathcal{V}_{t}$ and $\mathcal{V}_{t^{\prime}}$ in the personalized ranking $\rho_{i}\big(\{\mathcal{V}_{1},\mathcal{V}_{2},\ldots,\mathcal{V}_{|\mathcal{Q}|}\}\big)$ of visualizations for the individual user $i$ (from dataset $\mX_{\rm test}$)
such that $t<t^{\prime}$, then $\mathbcal{M}_i(\mathcal{V}_t) \geq \mathbcal{M}_i(\mathcal{V}_{t^{\prime}})$ holds by definition.
\end{Definition}

\section{Personalized Visualization Recommendation Framework} \label{sec:framework}
In this section, we present the framework for solving the personalized visualization recommendation problem from Section~\ref{sec:problem}.
In Section~\ref{sec:meta-feature-space}, we first describe the meta-feature learning approach for mapping user datasets to a shared universal meta-feature space where relationships between the corpus of tens of thousands of datasets can be automatically inferred and used for learning individual personalized models for each user.
Then Section~\ref{sec:shared-data-preference-space} introduces a graph model that captures the data preferences of users while Section~\ref{sec:shared-visualization-space} proposes graph models that naturally encode the visual preferences of users.
The personalized visualization recommendation models learned from the proposed graph representations are described in Section~\ref{sec:models}, while the visualization scoring and recommendation techniques are presented in Section~\ref{sec:inference}.

\subsection{Representing Datasets in a Universal Shared Meta-feature Space}\label{sec:meta-feature-space}
To learn from user datasets of different sizes, types, and characteristics, we first embed the attributes (columns) of each dataset $\mX \in \mathcal{X}_1 \cup \mathcal{X}_2 \cup \cdots \cup \mathcal{X}_n$ (from any user) in a shared $K$-dimensional meta-feature space.
This also enables the personalized visualization recommendation model to learn from users with similar data preferences.
Recall that each user $i \in [n]$ is associated with a set of datasets $\mathcal{X}_i=\{\mX_{i1}, \mX_{i2}, \ldots\}$.
\begin{Claim}\label{claim:users-share-some-datasets}
Let $\mathbcal{X} = \bigcup_{i=1}^n \mathcal{X}_i$ denote the set of all datasets. 
Then
\begin{align} \label{eq:users-share-some-datasets}
\sum_{i=1}^n |\mathcal{X}_i| \geq |\mathbcal{X}|
\end{align}
\end{Claim}\noindent
Hence, if $\sum_{i=1}^n |\mathcal{X}_i| = |\mathbcal{X}|$, then this implies that all users have completely different datasets (there does not exist any two users $i,j \in [n]$ that have a dataset in common).
Otherwise, if there exists two users that have at least one dataset in common with one another, then $\sum_{i=1}^n |\mathcal{X}_i| > |\mathbcal{X}|$.

In our personalized visualization recommendation problem (Sec.~\ref{sec:problem}), it is possible (and in many cases likely) that users are interested in completely different datasets.
In the worst case, every user has a completely disjoint set of datasets, and thus, the implicit and/or explicit user feedback regarding the attributes of interest to the users is also completely disjoint.
In such a case, the question then becomes how can we leverage the feedback from users like this, to better recommend attributes from 
different datasets that may be of interest to a new and/or previous user?
To do this, we need a general method that can derive a fixed-size embedding $\Psi(\vx) \in \RR^{K}$ of an attribute $\vx$ from any arbitrary dataset $\mX$ such that the $K$-dimensional embedding $\Psi(\vx)$ captures the important data characteristics and statistical properties of $\vx$, independent of the dataset and size of $\vx$.
Afterwards, given two attributes $\vx$ and $\vy$ from different datasets (\ie, $\mX$ and $\mY$)
and users, 
we can derive the similarity between $\vx$ and $\vy$.
Suppose there is implicit/explicit user feedback regarding an attribute $\vx$,
then given another arbitrary user $i$ interested in a new dataset $\mY$ (without any feedback on the attributes in $\mY$), then we can derive the similarity between $\vx$ and $\vy$, and if $\vy$ is similar to an attribute $\vx$ that was preferred by some user(s), then we can assign the attribute $\vy$ a higher probability (weight, score), despite that it doesn't yet have any user feedback.
Therefore, as discussed above, it is clear that this idea of transferring user feedback about attributes across different datasets is extremely powerful and fundamentally important for personalized visualization recommendation (especially when there is only limited sparse feedback available).
Moreover, the proposed idea above is also important when there is no feedback about an attribute in some dataset, or a completely new dataset of interest by a user.
This enables us to learn better personalized visualization recommendation models for individual users while requiring significantly less feedback.
\begin{Property}\label{prop:attr-similarity-epsilon}
Two attributes $\vx$ and $\vy$ are similar iff 
\begin{align}\label{eq:attr-similarity-epsilon}
s\langle \Psi(\vx), \Psi(\vy) \rangle > 1-\epsilon.
\end{align}
\end{Property}\noindent
where $s\langle \cdot,\cdot \rangle$ is the similarity function.
Notice that since almost surely $|\vx| \not= |\vy|$ (different sizes), then the similarity of $\vx$ and $\vy$ cannot be computed directly.
Therefore, we embed $\vx$ and $\vy$ into the same $K$-dimensional meta-feature space where there similarity can be computed directly as $s\langle \Psi(\vx), \Psi(\vy) \rangle$.

\begin{table*}
\centering
\caption{Meta-feature learning framework overview} 
\label{table:meta-feature-learning-framework}
\vspace{-2mm}
\small
\begin{tabular}{l l  l l  } 
\toprule\textsc{Framework Components}  & \textsc{Examples} &\\
\midrule
\textbf{1. Data representations} $\mathcal{G}$ & $\vx$, $\vp$, $g(\vx)$, $\ell_{b}\!(\vx)$ log-binning, ...  &  &  \\ 		
\textbf{2. Partitioning functions} $\Pi$ & Clustering, binning, quartiles, ...  &  & \\
\textbf{3. Meta-feature functions} $\psi$ & Statistical, information theoretic, ...  &  & \\ 
\textbf{4. Meta-embedding of meta-features} & $\argmin\limits_{\mH,\boldsymbol{\Sigma}, \mQ}\; \mathbb{D}_{\!\!\mathcal{L}}\big(\mM \| \mH\boldsymbol{\Sigma}\mQ^{\top}\big)$,\quad then 
$\hat{\vq} = \boldsymbol{\Sigma}^{-1}\mH^{\top}\hat{\vm}$  &  & \\ 
\bottomrule
\end{tabular}
\end{table*}

Attributes from different datasets are naturally of different sizes, types, and even from different domains.
Therefore as shown above, there is no way to compute similarity between them directly.
Instead, we propose to map each attribute from any arbitrary dataset into a shared $K$-dimensional space using meta-feature functions.
After every attribute is mapped into this $K$-dimensional meta-feature space, we can then compare their similarity directly.
In this work, we propose a meta-feature learning framework with four main components as shown in Table~\ref{table:meta-feature-learning-framework}.
Many of the framework components use the meta-feature functions denoted as $\psi$.
A meta-feature function is a function that maps an arbitrary vector to a value that captures a specific characteristic of the vector of values.
In this work, we leverage a large class of meta-feature functions formally defined in Table~\ref{table:meta-feature-functions}.
However, the framework is flexible and can leverage any arbitrary collection of meta-feature functions.
Notably, mapping every attribute from any dataset into a low-dimensional meta-feature space enables the model to capture and learn from the similarity between user preferred attributes in completely different datasets.

Let $\vx$ denote an attribute (column vector) from any arbitrary user dataset.
Then we may apply the collection of meta-features $\psi$ from Table~\ref{table:meta-feature-functions} directly to $\vx$ to obtain a low-dimensional representation of $\vx$ as $\psi(\vx)$.
In addition, we can also apply the meta-feature functions $\psi$ to various representations and transformation of $\vx$.
For instance, we can first derive the probability distribution $p(\vx)$ of $\vx$ such that $p(\vx)^{\top}\ve=1$, and then use the meta-feature functions $\psi$ over $p(\vx)$ to characterize this representation of $\vx$. 
We can also use the meta-feature functions to characterize other important representations and transformations of the attribute vector $\vx$ such as different scale-invariant and dimensionless representations of the data using different normalization functions $g_h(\cdot)$ over the attribute (column) vector $\vx$, and from each of these representations, we can apply the above meta-feature functions, \eg, $g_h(\vx) = \frac{\vx - \min(\vx)}{\max(\vx) - \min(\vx)}$, then $\psi(g_h(\vx))$.
More generally, let $\mathcal{G} = \{g_1, g_2, \ldots, g_{\ell}\}$ denote a set of data representation and transformation functions that can be applied over an attribute vector $\vx$ from any arbitrary user dataset. 
We first compute the meta-feature functions $\psi$ 
(\eg, from Table~\ref{table:meta-feature-functions}) 
over the $\ell$ different representations of the attribute vector $\vx$ given by the functions $\mathcal{G} = \{g_1, g_2, \ldots, g_{\ell}\}$ as follows:
\begin{equation}\label{eq:meta-feature-different-data-representations}
\psi(g_1(\vx)), \psi(g_2(\vx)), \ldots, \psi(g_{\ell}(\vx))
\end{equation}\noindent
Note that if $g \in \mathcal{G}$ is the identity function, then $\psi(g(\vx))=\psi(\vx)$.
In all cases, the meta-feature function $\psi$ maps a vector of arbitrary size to a fixed size  lower-dimensional vector.

For each of the different representation/transformation functions $\mathcal{G} = \{g_1,\ldots,g_{\ell}\}$ of the attribute vector $\vx$, we use a partitioning function $\Pi$ to group the different values into $k$ different subsets (\ie, partitions, clusters, bins).
Then we apply the meta-feature functions $\psi$ to each of the $k$ different groups as follows:
\begin{align}\label{eq:meta-features-of-partitions-single-partition-function}
\underbrace{\psi(\Pi_{1}(g_1(\vx))), \ldots, \psi(\Pi_{k}(g_1(\vx)))}_{g_1(\vx)}, 
\ldots,
\underbrace{\psi(\Pi_{1}(g_{\ell}(\vx))), \ldots, \psi(\Pi_{k}(g_{\ell}(\vx)))}_{g_{\ell}(\vx)}
\end{align}
where $\Pi_{k}$ denotes the $k$th partition of values from the partitioning function $\Pi$.
Note that to ensure every attribute is mapped to the same $K$-dimensional meta-feature space, we only need to fix the number of partitions $k$.
In Eq.~\ref{eq:meta-features-of-partitions-single-partition-function}, we show only a single partitioning function $\Pi$, however, multiple partitioning functions are used in this work and each is applied in a similar fashion as Eq.~\ref{eq:meta-features-of-partitions-single-partition-function}.
All the meta-features 
derived from Eq.~\ref{eq:meta-feature-different-data-representations} and Eq.~\ref{eq:meta-features-of-partitions-single-partition-function} 
are then concatenated into a single vector of meta-features describing the characteristics of the attribute $\vx$.
More formally, the meta-feature function $\Psi : \vx \to \RR^{K}$ that combines the different components from the framework (in Table~\ref{table:meta-feature-learning-framework}) is defined as
\begin{align}\label{eq:meta-feature-vector}
\Psi(\vx) = \big[
\psi(g_1\!(\vx)) \cdots \psi(g_{\ell}(\vx))
&\cdots
\psi(\Pi_{1}(g_1\!(\vx))) \cdots \psi(\Pi_{k}(g_1\!(\vx)))
\\ \nonumber
&\cdots 
\psi(\Pi_{1}(g_{\ell}(\vx))) \cdots \psi(\Pi_{k}(g_{\ell}(\vx)))
\big] 
\end{align}\noindent
The resulting $\Psi(\vx)$ is a $K$-dimensional meta-feature vector for attribute $\vx$.
Our approach is agnostic to the precise meta-feature functions used, and is flexible for use with any alternative set of meta-feature functions (Table~\ref{table:meta-feature-functions}).

Let $\mathbcal{X} = \cup_{i=1}^n \mathcal{X}_i$ denote the set of dataset matrices across all $n$ users. 
Given an arbitrary dataset matrix $\mX \in \mathbcal{X}$ (which can be shared among multiple users), let $\Psi(\mX) \in \RR^{K \times |\mX|}$ be the resulting meta-feature matrix obtained by
applying $\Psi$ independently to each of the $|\mX|$ attributes (columns) of $\mX$.
Then, we can derive the overall meta-feature matrix $\mM$ as
\begin{align}\label{eq:meta-feature-matrix-over-all-users-and-datasets}
\mM = \bigoplus_{\mX \in \mathbcal{X}} \Psi(\mX)
\end{align}
where $\bigoplus$ is the concatenation operator, \ie, $\Psi(\vx) \oplus \Psi(\vy) = [\Psi(\vx)\, \Psi(\vy)] \in \RR^{K \times 2}$.
Note that Eq.~\ref{eq:meta-feature-matrix-over-all-users-and-datasets} is not equivalent to $\bigoplus_{i=1}^{n} \bigoplus_{\mX \in \mathcal{X}_i} \Psi(\mX)$ since any two users $i,j \in [n]$ can share one or more datasets.
With slight abuse of notation, let $d = |\mathbcal{X}|$ and $\mathbcal{X} = \{\mX_1,\ldots,\mX_d\}$, then $\mM = \Psi(\{\mX_1, ,\ldots,\mX_d\})$ where $\mM$ is a $K \times (|\mX_1|+\cdots+|\mX_d|)$ matrix.

\begin{figure*}[t!]
\centering
\subfigure{
\includegraphics[width=0.45\linewidth]{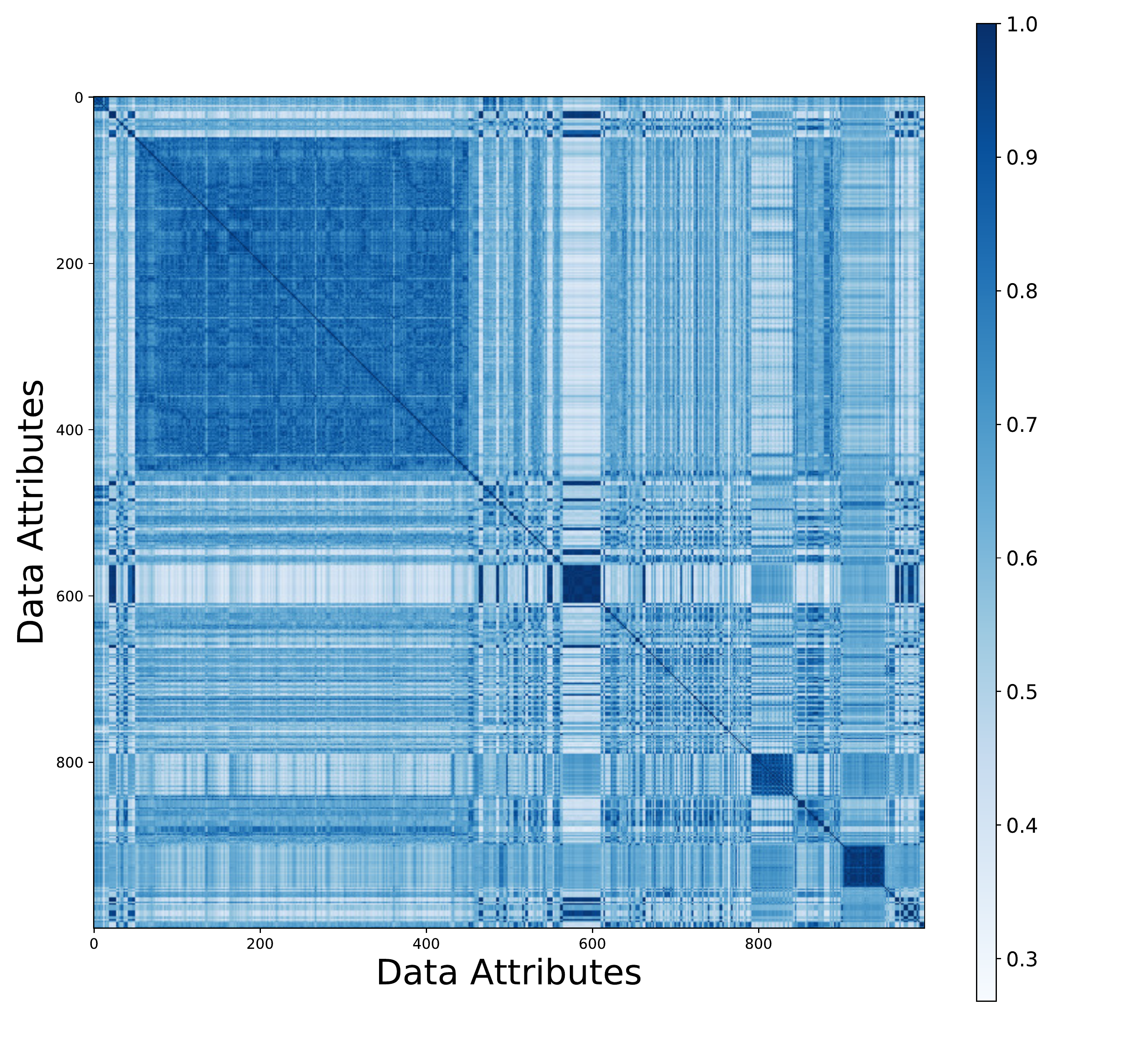}
\label{fig:attr-metaFeat-sim}
}
\hspace{4mm}
\subfigure{
\includegraphics[width=0.45\linewidth]{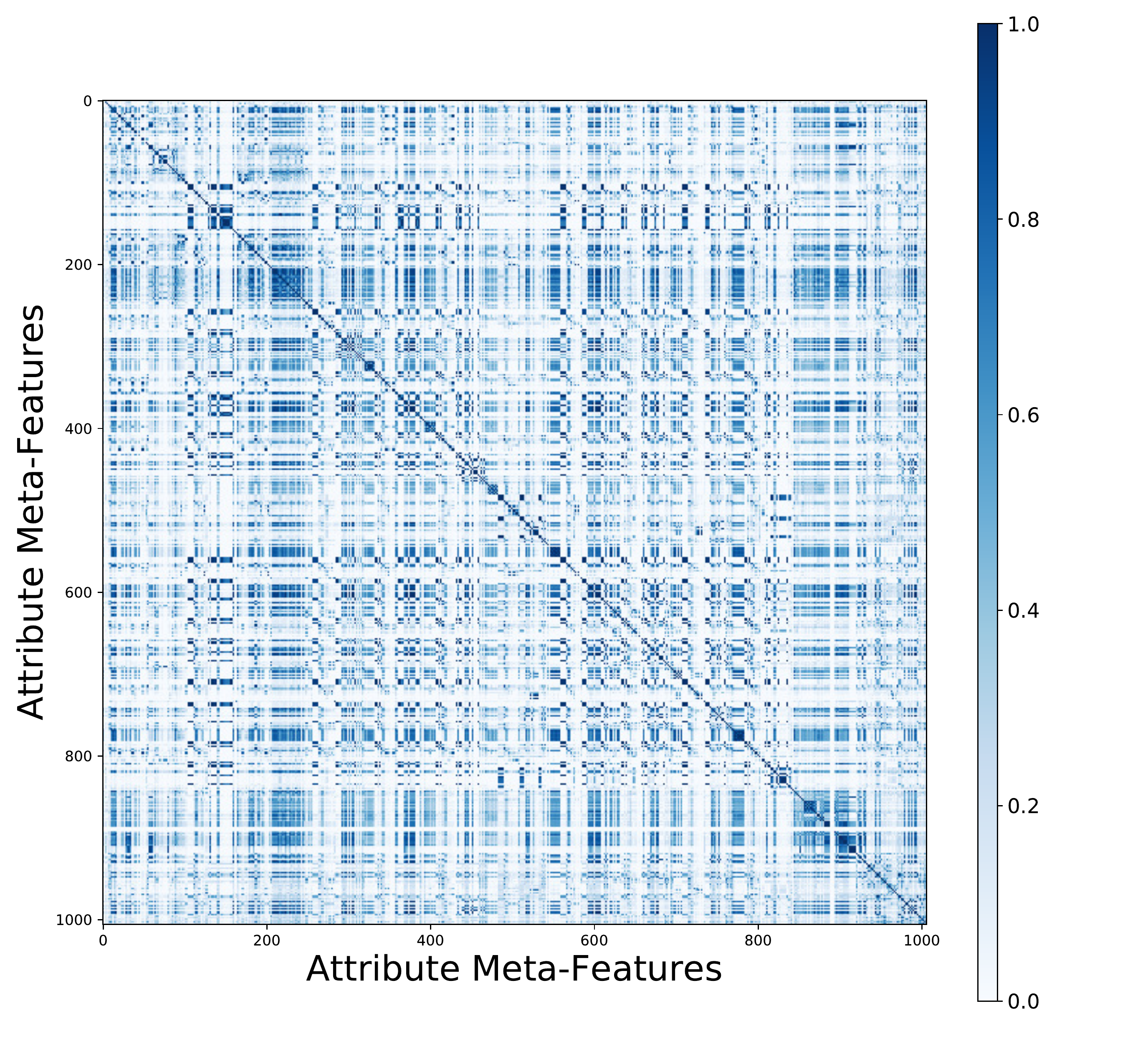}
\label{fig:metaFeat-sim}
}
\vspace{-1mm}
\caption{
Similarity of attributes across different datasets using the attribute embeddings in the universal meta-feature space.
(a) uses the first 1000 attributes across different datasets and takes the cosine similarity between each pair of attributes with respect to their fixed $k$-dimensional meta-feature vectors.
(b) shows the cosine similarity between the attribute meta-features used to characterize the different attributes.
See text for discussion.
}
\label{fig:attr-sim}
\end{figure*}

In Figure~\ref{fig:attr-sim}, we investigate the similarity of attributes across different datasets in the personalized visualization corpus (Section~\ref{sec:dataset}), and observe two important findings.
First, Figure~\ref{fig:attr-metaFeat-sim} indicates that attributes across different datasets may be similar to one another and the latent relationships between the attributes can benefit learning personalized visualization recommendation models, especially for users with very few or even no visualization feedback.
Second, the meta-features used to characterize attributes from any arbitrary dataset are diverse and  fundamentally different from one another as shown in Figure~\ref{fig:metaFeat-sim}.
This finding is important and validates the proposed meta-feature learning framework since the meta-features must be able to capture the fundamental patterns and characteristics for a dataset from any arbitrary domain.

\begin{table*}
\centering
\caption{Summary of attribute meta-feature functions. Let $\vx$ denote an arbitrary attribute (variable, column) vector and $\pi(\vx)$ is the sorted vector.} 
\label{table:meta-feature-functions}
\vspace{-2mm}
\small
\def\arraystretch{0.9}
\begin{tabular}{l l  l @{}H HHH  @{}} 
\toprule
\textbf{Function Name}  &
\textbf{Equation} &
\textbf{Rationale} & 
\\
\midrule
Num. instances  &  $\abs{\vx}$ & Speed, Scalability         & $p/\abs{\vx}$, $log(\abs{\vx})$, $log(\abs{\vx}/p)$ \\ 
Num. missing values &  $s$ & Imputation effects         &  \\
Frac. of missing values &  $\nicefrac{\abs{\vx} - s}{\abs{\vx}}$ & Imputation effects  &  \\
Num. nonzeros & $\mathtt{nnz}(\vx)$ & Imputation effects & \\
Num. unique values & $\mathsf{card}(\vx)$ & Imputation effects & \\
Density & $\nicefrac{\mathsf{nnz}(\vx)}{\abs{\vx}}$ & Imputation effects & \\
\midrule
$Q_1$, $Q_3$  & median of the $\abs{\vx}/2$ smallest (largest) values & $-$ & \\
IQR  & $Q_3 - Q_1$ & $-$ & \\
Outlier LB $\alpha \in \{1.5,3\}$ & $\sum_{i} \mathbb{I}(x_i < Q_1-\alpha IQR)$ & Data noisiness & \\
Outlier UB $\alpha \in \{1.5,3\}$ & $\sum_{i} \mathbb{I}(x_i > Q_3+\alpha IQR)$ & Data noisiness & \\
Total outliers $\alpha \in \{1.5,3\}$ & 
$\sum_{i} \mathbb{I}(x_i < Q_1-\alpha IQR) + \sum_{i} \mathbb{I}(x_i > Q_3+\alpha IQR)$ 
& Data noisiness & \\
($\alpha$std) outliers $\alpha \in \{2,3\}$        &  $\mu_{\vx} \pm \alpha \sigma_{\vx}$ & Data noisiness             & $o/\abs{\vx}$, lb, ub, total \\
\midrule
Spearman ($\rho$, p-val)  & $\mathsf{spearman}(\vx, \pi(\vx))$ & Sequential & \\
Kendall ($\tau$, p-val) & $\mathsf{kendall}(\vx, \pi(\vx))$ & Sequential & \\
Pearson ($r$, p-val) & $\mathsf{pearson}(\vx, \pi(\vx))$ & Sequential & \\	
\midrule
Min, max & $\min(\vx)$, $\max(\vx)$ &  $-$ \\
Range  & $\max(\vx) - \min(\vx)$ & Attribute normality \\
Median  & $\mathsf{med}(\vx)$ &  Attribute normality \\
Geometric Mean  & $\abs{\vx}^{-1} \prod_i x_i $ & Attribute normality & \\
Harmonic Mean & $\abs{\vx} / \sum_i \frac{1}{x_i}$ & Attribute normality & \\
Mean, Stdev, Variance  & $\mu_{\vx}$, $\sigma_{\vx}$, $\sigma^2_{\vx}$ & Attribute normality & \\
Skewness  & $\nicefrac{\mathbb{E}(\vx - \mu_{\vx})^3}{\sigma^3_{\vx}}$ & Attribute normality & \\
Kurtosis  & $\nicefrac{\mathbb{E}(\vx - \mu_{\vx})^4}{\sigma^4_{\vx}}$ & Attribute normality &
Fisher/Pearson, and bias/unbiased. \\
HyperSkewness  & $\nicefrac{\mathbb{E}(\vx - \mu_{\vx})^5}{\sigma^5_{\vx}}$ & Attribute normality & \\
Moments [6-10]  & $-$ & Attribute normality & \\
k-statistic [3-4]  & $-$ & Attribute normality &  \\
\midrule
Quartile Dispersion Coeff. & $\frac{Q_3-Q_1}{Q_3+Q_1}$ & Dispersion & \\
Median Absolute Deviation & $\mathsf{med}(\abs{\vx - \mathsf{med}(\vx)})$ & Dispersion & \\
Avg. Absolute Deviation & $\frac{1}{\abs{\vx}} \ve^T\!\abs{\vx - \mu_{\vx}} $ & Dispersion & \\
Coeff. of Variation & $\nicefrac{\sigma_{\vx}}{\mu_{\vx}}$ & Dispersion & \\
Efficiency ratio & $\nicefrac{\sigma^2_{\vx}}{\mu^2_{\vx}}$ & Dispersion & \\
Variance-to-mean ratio & $\nicefrac{\sigma^2_{\vx}}{\mu_{\vx}}$ & Dispersion & \\
\midrule
Signal-to-noise ratio (SNR)  & $\nicefrac{\mu^2_{\vx}}{\sigma^2_{\vx}}$ & Noisiness of data  & \\
Entropy & $H(\vx) = -\sum_{i} \; x_i \log x_i$ & Attribute Informativeness  & \\
Norm. entropy & $\nicefrac{H(\vx)}{\log_2 \abs{\vx}}$ & Attribute Informativeness  & \\
Gini coefficient & $-$ & Attribute Informativeness & \\
\midrule
Quartile max gap & $\max (Q_{i+1} - Q_{i}) $ & Dispersion & \\
Centroid max gap & $\max_{ij} |c_{i} - c_{j}| $ & Dispersion & \\
\midrule
Histogram prob. dist. & $\vp_h = \frac{\vh}{\vh^T\ve}$ (with fixed \# of bins) & - & \\
\midrule
\text{Landmarker(4-Means)} & \multicolumn{3}{l}{
(i) sum of squared dist.,
(ii) mean silhouette coeff., 
(iii) num. of iterations
} \\
\bottomrule
\end{tabular}
\end{table*}

\subsubsection{Meta-Embedding of Meta-Features} \label{sec:meta-embedding-of-meta-features}
We can derive an embedding using the current meta-feature matrix $\mM$.
Note this meta-feature matrix may contain all meta-features across all previous datasets or simply the meta-features of a single dataset. However, the more datasets, the better the meta-embedding of the meta-feature will reveal the important latent structures between the meta-features.
We learn the latent structure in the meta-feature matrix $\mM$ by solving\footnote{Assume \emph{w.l.o.g.} that columns of $\mM$ and the meta-features of a new attribute $\hat{\vm}$ are normalized to length 1.}
\begin{equation} \label{eq:meta-feature-embedding}
\argmin\limits_{\mH,\boldsymbol{\Sigma}, \mQ}\; \mathbb{D}_{\!\!\mathcal{L}}\big(\mM \| \mH\boldsymbol{\Sigma}\mQ^{\top}\big)
\end{equation}
\noindent
Given meta-features for a new attribute $\hat{\vm}$ in another arbitrary unseen dataset, we use the latent low-rank meta-embedding matrices to map the meta-feature vector $\hat{\vm} \in \RR^{k}$ into the low-rank meta-embedding space as
\begin{equation}
\hat{\vq} = \boldsymbol{\Sigma}^{-1}\mH^{\top}\hat{\vm}
\end{equation}
Hence, the meta-feature vector $\hat{\vm}$ of a new previously unseen attribute is mapped into the same meta-embedding space $\hat{\vq}$.
The resulting meta-embedding of the meta-features of the new attribute can then be concatenated onto the meta-feature vector.
This has several important advantages.
First, using the proposed meta-feature learning framework shown in Table~\ref{table:meta-feature-learning-framework} results in hundreds or thousands of meta-features for a single attribute. 
Many of these meta-features may not be important for a specific attribute, while a few meta-features may be crucial in describing the attribute and its data characteristics.
Therefore, the meta-embedding of the meta-features can be viewed as a noise reduction step that essentially removes redundant or noisy signals from the data while preserving the most important signals that describe the fundamental direction and characteristics of the data.
Second, the meta-embedding of the meta-features reveals the latent structure and relationships in the meta-features.
This step can also be viewed as a type of landmark feature since we solve a learning problem to find a low-rank approximation of $\mM$ such that $\mM \approx \mH\boldsymbol{\Sigma}\mQ^{\top}$.

However, we include it as a different component of the meta-feature learning framework in Table~\ref{table:meta-feature-learning-framework} since instead of concatenating the meta-embedding of the meta-features for a attribute, we can also use it directly by replacing it with the meta-feature vector.
This is especially important when there is a large number of datasets (\eg, more than 100K datasets with millions of attributes in total) for learning.
For instance, if there are 2.3M attributes (see 100K dataset in Table~\ref{table:dataset-statistics}), and each attribute is encoded with a dense $K=1006$ dimensional meta-feature vector, then $\mM$ has $1006 \times 2,300,000$ values that need to be stored, which use 18.5GB space (assuming 8 bytes per value).
However, if we use the meta-embedding of the meta-features with $K=10$, then $\mM$ takes about 200MB (0.18GB) of space.

\subsection{Learning from User-level Data Preferences Across Different Datasets}\label{sec:shared-data-preference-space}
Given users $i$ and $j$ that provide feedback on the attributes of interest from two completely different datasets, how can we leverage the user feedback (data preferences) despite it being across different datasets without any shared attributes?
To address this important problem, we propose a novel representation and model that naturally enables the transfer of user-level data preferences across different datasets to improve predictive performance, recommendations, and reduce data sparsity.
The across dataset transfer learning of user-level data preferences becomes possible
due to the proposed representation and model for personalized visualization recommendation.

We now introduce the novel user-level data preference graph model for personalized visualization recommendation that naturally enables across-dataset and across-user transfer learning of preferences.
This model encodes users, their interactions with attributes (columns/variables from any arbitrary dataset) and the meta-features of the attributes. 
This new representation enables us to learn from user-level data preferences across different datasets and users, and therefore very important for personalized visualization recommendation systems.
In particular, we first derive the following user-by-attribute preference matrix $\mA$ as follows:
\begin{align}\label{eq:user-by-attr-matrix}
\mA = \big[\mA\big]_{ij} = \!\text{\# of times user $i$ clicked 
(a visualization with) 
attribute $j$ 
}
\end{align}\noindent
In terms of the implicit or explicit user ``action'' encoded by $A_{ij}$, it could be an implicit user action such as when a user clicks or hovers-over a specific attribute $j$ or when a user clicks or hovers-over a visualization that uses attribute $j$ in it.
Similarly, $A_{ij}$ can encode an explicit user action/feedback such as the attribute $j$ that a user explicitly liked (independent of a visualization), or more generally, the attribute $j$ used in a \emph{visualization} that a user explicitly liked, or added-to-their dashboard, and so on.
In other words, there are two different types of explicit and implicit user feedback about attributes, notably, user feedback regarding a visualization that used an attribute $j$ (whether the user action is a click, hover, added-to-dashboard, etc), or more directly, whether a user liked or clicked on an attribute in the dataset directly via some UI.

Given $\mA$ defined in Eq.~\ref{eq:user-by-attr-matrix}, we are able to learn from two or more users that have at least one attribute preference in common.
More precisely, $\mA_{i,:}^{\top} \mA_{j,:} > 0$ for two arbitrary users $i$ and $j$, which implies two users $i$ and $j$ share a dataset of interest, \emph{and} have preferred at least one of the same attributes in that dataset.
Unfortunately, finding two users that satisfy the above constraint is often unlikely.
Therefore, we need to add another representation to $\mA$ that creates meaningful connections between attributes in different datasets based on their similarity.
In particular, we do this by leveraging the meta-feature matrix $\mM$ from Section~\ref{sec:meta-feature-space} derives by mapping every attribute in a user-specific dataset to a k-dimensional meta-feature vector. 
This defines a universal meta-feature space that is shared among the attributes in any arbitrary dataset, and therefore allowing the learning of connections between users and their preferred attributes in completely different datasets. 
This new component is very important, since without it, we have no way to learn from other users (and across different datasets), since each user has its own datasets, and thus has their own set of visualizations (where each visualization consists of a set of design choices and data choices) that are not shared by any other users. 

\begin{figure}[h!]
\vspace{-1.6mm}
\centering
\includegraphics[width=0.9\linewidth]{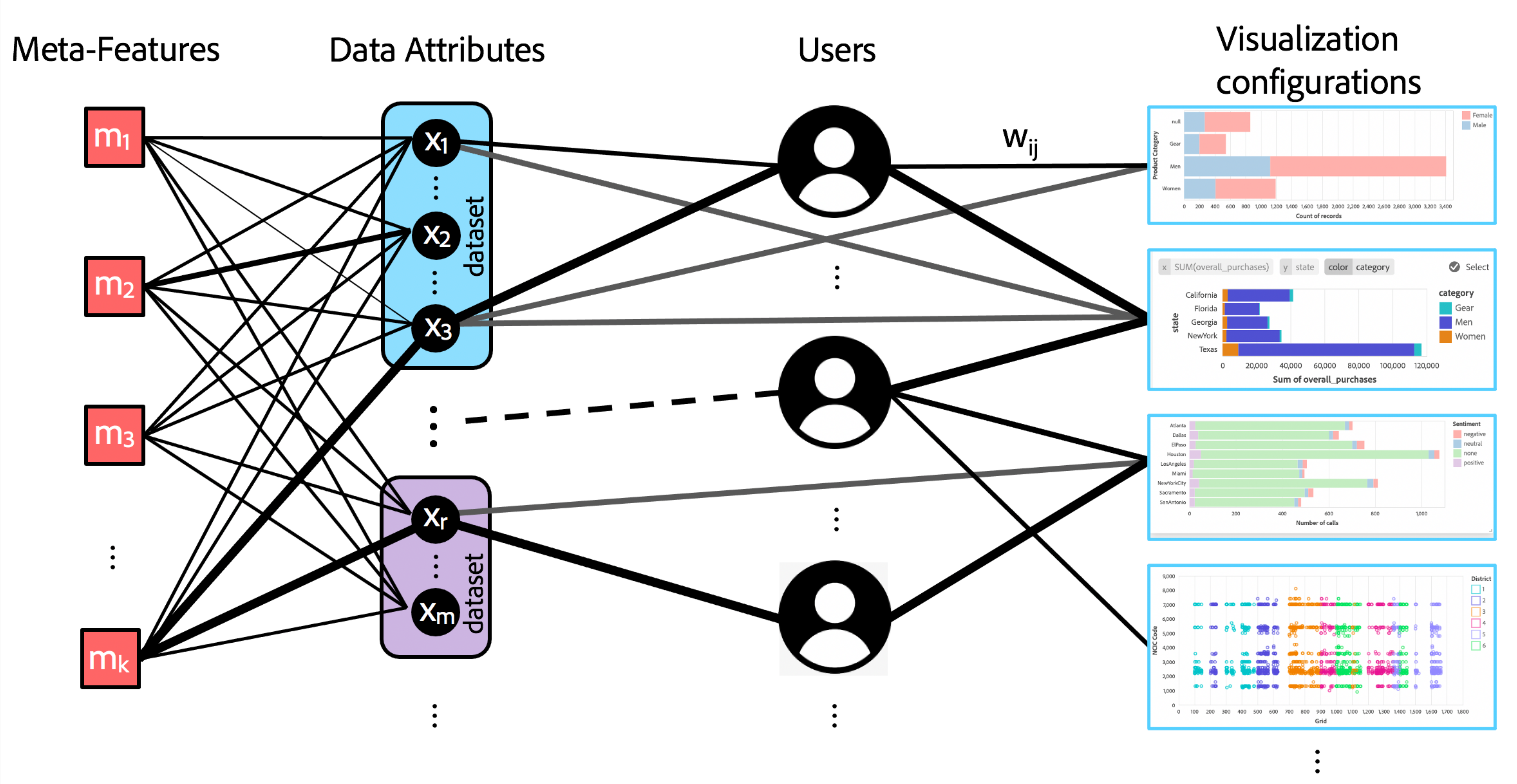}
\vspace{-3.6mm}
\caption{Overview of the proposed graph model for personalized visualization recommendation.
Links between meta-features and data attributes represent the meta-feature matrix $\mM$ from Section~\ref{sec:meta-feature-space} whereas links between users and their preferred data attributes represent $\mA$ (Section~\ref{sec:shared-data-preference-space}). 
Both of these capture the user-level data preferences across different datasets. 
Links between users and their visualization configurations represent $\mA$ and capture the user-level visual preferences.
Finally, links between visualization configurations and data attributes represent $\mD$.
Note that all links are weighted; and the data attribute/column nodes of a specific dataset are grouped together.
}
\label{fig:personalized-vis-rec-graph-model}
\vspace{-2.5mm}
\end{figure}

\subsection{Learning from \emph{User-level Visual Preferences} Across Different Datasets} \label{sec:shared-visualization-space}
Visualizations naturally consist of data and visual design choices.
The dependence of data in a visualization means that visualizations generated for one dataset will be completely different from the visualizations generated by any other dataset.
This is problematic if we want to develop a personalized visualization recommendation system that can learn from the \emph{visual preferences} of users despite that the users may not have preferred any visualizations from the same datasets.
This is important since visualizations are fundamentally tied to the dataset, and each user may have their own set of datasets that are not shared by any other user.
Moreover, even if two users had a dataset in common, the probability that the users prefer the same visualization is very small (almost surely zero) due to the exponential space of visualizations that may arise from a single dataset.

To overcome these issues, we introduce a dataset independent notion of a visualization called a visualization configuration. 
Using this notion, we propose a novel graph representation that enables us to learn from the \emph{visual preferences} of users despite that they may not have preferred any visualizations from the same datasets.
In particular, to learn from visualizations across different datasets and users, we introduce a dataset independent notion called a visualization configuration that removes the dataset dependencies of  visualizations, enabling us to capture the general user-level visual preferences of users, independent of the dataset of interest.
A visualization configuration is an abstraction of a visualization where instead of mapping specific data attributes to specific design choices of the visualization (\eg, x, y, color, etc.), we replace them with their general type (\eg, numerical, categorical, temporal, ...) or other general property or set of properties that generalize across the different datasets.
Most importantly, it is by replacing the data-specific design choices with their general type or set of general properties that enables us to capture and learn from these visualization-configurations.
More formally, 
\begin{Definition}[Visual Configuration]
\label{def:vis-config}
Given a visualization $\mathcal{V}$ consisting of a set of design choices and data (attributes) associated to a subset of the design choices.
For every design choice such as chart-type, there is a set of possible options.
Other design choices such as color can also be associated to a set of options, \eg, static color definitions, or color map for specific data attributes.
Let $\mathcal{T} : \vx \to \mathcal{P}$ be a function that maps an attribute $\vx$ of some arbitrary dataset $\mX \in \mathbcal{X}$ to a property $P$ that generalizes across any arbitrary dataset, and therefore, is independent of the specific dataset.
Hence, given attributes $\vx$ and $\vy$ from two different datasets, then it is possible that $\mathcal{T}(\vx) = P$ and $\mathcal{T}(\vy) = P$. 
A visualization configuration is defined as an abstraction of a visualization where 
every design choice of a visualization that is bound to a data attribute is replaced with a general property of the attribute $\mathcal{T}(\vx) = P$.
\end{Definition}\noindent

\begin{figure}[t!]
\centering
\includegraphics[width=0.9\linewidth]{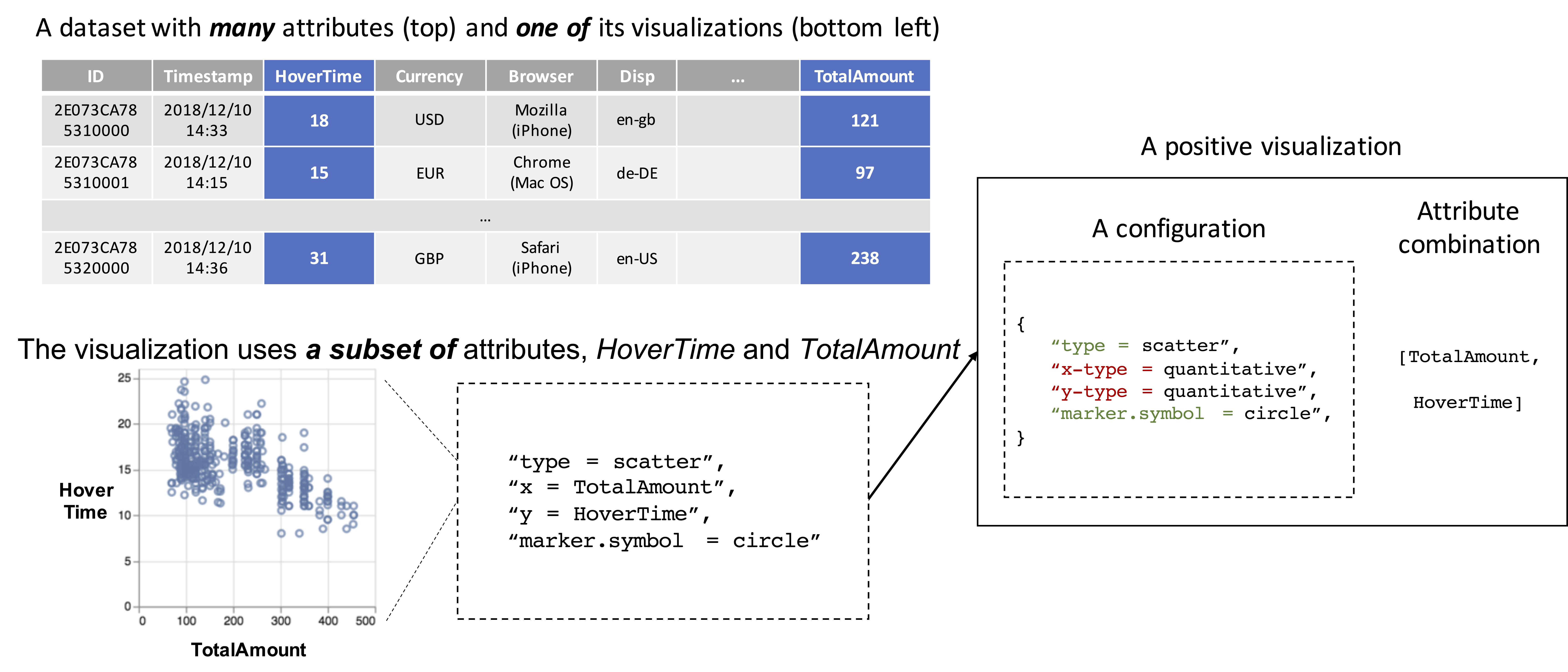}
\vspace{-3mm}
\caption{Visualization to Data Independent Visual Configuration.
A visualization consists of the data along with a set of design choices, and thus are dataset dependent.
In this work, we propose the notion of a visual configuration that removes the data dependency of a visualization while capturing the visual design choices.
Notably, visual configurations naturally generalize across datasets, since they are independent of the dataset, and thus unlike visualizations, a visual configuration \emph{can} be shared among users that use entirely different datasets.
The visualization shown in the above toy example uses only two data attributes from the dataset.
Note actual visual configurations have many other design choices which are not shown in the toy example above for simplicity.
}
\label{fig:vis-to-config}
\end{figure}

\begin{Claim}
There exists $\vx$ and $\vy$ from different datasets such that 
$\mathcal{T}(\vx) = P$ and $\mathcal{T}(\vy) = P$ hold.
\end{Claim}\noindent
The size of the space of visualization configurations is large since visualization configurations come from all possible combinations of design choices and their values such as, 
\begin{itemize}[label={}]
\item \textbf{chart-type:} bar, scatter, ...
\item \textbf{x-type:} quantitative, nominal, ordinal, temporal, ..., none
\item \textbf{y-type:} quantitative, nominal, ordinal, temporal, ..., none
\item \textbf{color:} red, green, blue, ...
\item \textbf{size:} 1pt, 2pt, ...
\item \textbf{x-aggregate:} sum, mean, bin, ..., none
\item \textbf{y-aggregate:} sum, mean, bin, ..., none
\item \textbf{...} 
\end{itemize} 
A visualization configuration \emph{and} the attributes selected is everything necessary to generate a visualization.
\noindent
In Figure~\ref{fig:vis-to-config}, we provide a toy example showing the process of extracting a data-independent visual-configuration from a visualization.
Using the notion of a visualization configuration (Definition~\ref{def:vis-config}), we can now introduce a model that captures the visualization preferences of users while ensuring that the visual preferences are not tied to specific datasets. 
In particular, we define the visual preference matrix $\mC$ as follows:
\begin{equation} \label{eq:user-by-vis-config-matrix}
\mC = \big[\mC\big]_{ij} = \text{\# of times user i clicked 
visualization configuration j}
\end{equation}\noindent
Note that clicked is simply one such example.
Other possibilities of defining $\mC$ (or other similar visual preference matrices) include 
$C_{ij} = \#$ of times user $i$ performed $\text{action} \in \{\text{clicked}$, $\text{hovered}$, $\text{liked}$, $\text{added-to-dashboard}\}$ visualization configuration $j$.
From our proposed graph model shown in Eq.~\ref{eq:user-by-vis-config-matrix}, 
we can directly learn from user-level visual preferences across different datasets.
This novel user-level visual preference graph model for visualization recommendation encodes users and their visual-configurations.
Since each visual-configuration node represents a set of design choices that are by definition not tied to a user-specific dataset, then the model can use this user-level visual graph to infer and make connections between other similar visual-configurations likely to be of interest to that user.
This new graph model is critical since it allows the learning component to learn from user-level visual preferences (which are visual-configurations) across the different datasets and users. Without this novel component, there would be no way to learn from other users visual preferences (sets of design choices).

The novel notion of a visualization configuration that removes the dataset dependencies of a visualization enabling us to model and learn from the visual preferences of users across different datasets. 
We introduce the notion of a visualization-configuration, which is an abstraction of a visualization. 
For instance, suppose we have a json encoding of the actual visualization, that is the design choices + the actual attributes and their data used in the visualization (hence, using this json, we can create the visualization precisely). 
Recall that this is not very useful for personalized visualization recommendation since the visualization is clearly tied to the specific dataset used by a single user. 
Hence, if we used visualizations directly, then the optimization method used to optimize an arbitrary objective function to obtain the embeddings for inference would not be able to use other user preferences, since they would also be for visualizations tied to other datasets. 
To overcome this issue, we propose the novel notion of a visualization-configuration that removes the data-dependency. 
In particular, given a visualization which includes the design choices + data choices (e.g., data used for the x, y, color attributes), we derive a visualization-configuration from it by replacing the data (attributes) and data attribute names by general properties that are dataset-independent. 
For instance, in this work, we have used the type of the attribute (e.g., categorical, real-valued, etc.), but we can also use any other general property of the data as well. 
Most importantly, this new abstraction enables us to learn from users and their visual preferences (design choices), despite that these visual preferences are for visualizations generated for a completely different dataset. 
This is because we carefully designed the notion of a visualization-configuration to generalize across datasets. 
In other words, the proposed notion of visualization-configuration are independent of the dataset at hand, and therefore can be shared among users. 
Notice that traditional recommender systems used in movie or product recommendation are comparatively simple, since these systems assume a single universal dataset (set of movies, set of items/products) that all users share and have feedback about. 
However, none of these simple assumptions hold in the case of visualization recommendation, and therefore we had to develop and propose these new notions and models for learning.

Notice the matrix $\mC$ encodes the data-independent visual preferences of each user, which is very important.
However, this representation does not capture how the visual configurations map to the actual data preferences of the users (attributes and general meta-features that characterize the attributes).
Therefore, we also introduce another representation to encode these important associations.
In particular, we encode the attributes associated with each visual-configurations as,
\begin{align}\label{eq:vis-config-by-attribute-matrix}
\mD = \big[\mD\big]_{kt} = \text{\# of times attribute $k$ was used in visual-configuration $t$ clicked by some user}
\end{align}
As an example, given a relevant visualization $\mathcal{V}=(\mX_{ij}^{(k)}, C_{t}) \in \mathbcal{V}_{ij}$ of user $i \in [n]$ for dataset $\mX_{ij}$ with attributes $\mX_{ij}^{(k)}=[\vx_p\, \vx_q]$ and visual-configuration $C_{t} \in \mathbcal{C}$, 
we set $D_{pt} = D_{pt} + 1$ and $D_{qt} = D_{qt} + 1$.
We repeat this for all relevant visualizations of each user.
In Figure~\ref{fig:personalized-vis-rec-graph-model}, we provide an overview of the proposed personalized visualization recommendation graph model.

\subsection{Models for Personalized Visualization Recommendation} \label{sec:models}
We first introduce the PVisRec model that uses the learned meta-feature matrix $\mM$ from Section~\ref{sec:meta-feature-space} 
and all the graph representations proposed in Section~\ref{sec:shared-data-preference-space} for capturing the shared data preferences between users despite using completely different datasets 
along with the graph representations from Section~\ref{sec:shared-visualization-space} that capture the  visual preferences of users across all datasets in the corpus.
Then we discuss two variants of PVisRec that are investigated later in Section~\ref{sec:exp}.

\subsubsection{PVisRec:} \label{sec:PVisRec}
Given the sparse user by attribute adjacency matrix $\mA \in \RR^{n \times m}$, 
dense meta-feature by attribute matrix $\mM \in \RR^{k \times m}$, 
sparse user by visual-configuration adjacency matrix $\mC \in \RR^{n \times h}$, and
sparse attribute by visual-configuration adjacency matrix $\mD \in \RR^{m \times h}$,
the goal is to find the rank-$d$ embedding matrices $\mU$, $\mV$, $\mZ$, and $\mY$ that minimize the following objective function:
\begin{align} \label{eq:obj-func-full}
f(\mU, \mV, \mZ, \mY) = 
\|\mA - \mU\mV^{\top}\|^2 + 
\|\mM - \mY\mV^{\top}\|^2 +
\|\mC - \mU\mZ^{\top}\|^2 +
\|\mD - \mV\mZ^{\top}\|^2 
\end{align}
where $\mU \in \RR^{n \times d}$, $\mV \in \RR^{m \times d}$, $\mZ \in \RR^{h \times d}$, $\mY \in \RR^{k \times d}$ are low-rank $d$-dimensional embeddings of the users, attributes (across all datasets), visual-configurations, and meta-features. 
Further, the formulation above uses squared error, though other loss functions can also be used (e.g., Bregman divergences)~\cite{singh2008relational}.
We can solve Eq.~\ref{eq:obj-func-full} by computing the gradient and then using a first-order optimization method~\cite{schenker2021optimization}.
Afterwards, we have
\begin{align} 
\mA &\approx \mA^{\prime} = \mU\mV^{\top} = \sum_{r=1}^d \vu_r \vv_r^{\top} \\
\mM &\approx \mM^{\prime} = \mY\mV^{\top} = \sum_{r=1}^d \vy_r \vv_r^{\top} \\
\mC &\approx \mC^{\prime} = \mU\mZ^{\top} = \sum_{r=1}^d \vu_r \vz_r^{\top} \\
\mD &\approx \mD^{\prime} = \mV\mZ^{\top} = \sum_{r=1}^d \vv_r \vz_r^{\top}
\end{align}\noindent
Solving Eq.~\ref{eq:obj-func-full} corresponds to the \emph{PVisRec} model investigated later in Section~\ref{sec:exp}.
We also investigate a few different variants of the PVisRec model from Eq.~\ref{eq:obj-func-full} later in Section~\ref{sec:exp}.
In particular, the model variants of PVisRec use only a subset of the graph representations $\{\mA, \mC, \mD\}$ and/or dense meta-feature matrix $\mM$ introduced previously in Section~\ref{sec:meta-feature-space}-\ref{sec:shared-visualization-space}.

\subsubsection{PVisRec ($\mA$,$\mC$,$\mM$ only):}
Given the user by attribute matrix $\mA \in \RR^{n \times m}$, 
meta-feature by attribute matrix $\mM \in \RR^{k \times m}$, and 
user by visual-configuration matrix $\mC \in \RR^{n \times h}$, 
the goal is to find the rank-$d$ embedding matrices $\mU$, $\mV$, $\mZ$, and $\mY$ that minimize the following objective function:
\begin{align} \label{eq:obj-func-simple-A-C-M}
f(\mU, \mV, \mZ, \mY) = 
\|\mA - \mU\mV^{\top}\|^2 + 
\|\mM - \mY\mV^{\top}\|^2 +
\|\mC - \mU\mZ^{\top}\|^2 
\end{align}

\subsubsection{PVisRec ($\mA$,$\mC$,$\mD$ only)}
Besides Eq.~\ref{eq:obj-func-simple-A-C-M} that uses only $\mA$, $\mM$, and $\mC$, we also investigate another personalized visualization recommendation model that uses $\mA$, $\mC$, and $\mD$ (without meta-features).
More formally, given $\mA$, $\mC$, and $\mD$, then the problem is to learn low-dimensional rank-$d$ embedding matrices $\mU$, $\mV$, and $\mZ$ that minimize the following:
\begin{align} \label{eq:obj-func-simple-A-C-D}
f(\mU, \mV, \mZ) = 
\|\mA - \mU\mV^{\top}\|^2 + 
\|\mC - \mU\mZ^{\top}\|^2 +
\|\mD - \mV\mZ^{\top}\|^2 
\end{align}
\noindent
In this work, we used an ALS-based optimizer to solve Eq.~\ref{eq:obj-func-full} and the simpler variants shown in Eq.~\ref{eq:obj-func-simple-A-C-M} and Eq.~\ref{eq:obj-func-simple-A-C-D}.
However, we can also leverage 
a variety of different optimization schemes including cyclic/block coordinate descent~\cite{kim2014algorithms,pcmf-snam16}, stochastic gradient descent~\cite{yun2014nomad,oh2015fast}, among others~\cite{singh2008relational,bouchard2013convex,choi2019s3,balasubramaniam2020column,schenker2021optimization}.

\subsection{Inferring Personalized Visualization Recommendations for Individual Users}
\label{sec:inference}
We first discuss using the personalized visualization recommendation model for recommending attributes to users as well as visual-configurations.
Then we discuss the fundamentally more challenging task of personalized visualization recommendation.

\subsubsection{Personalized Attribute Recommendation}
The ranking of attributes for user $i$ is induced by $\mU_{i,:}\mV^{\top}$ where $\mU_{i,:}$ is the embedding of user $i$.
Let $\pi_1(\mU_{i,:}\mV^{\top})$ denote the largest attribute weight for user $i$.
Therefore, the top-$k$ attribute weights for user $i$ are denoted as:
\[
\pi_1(\mU_{i,:}\mV^{\top}), \pi_2(\mU_{i,:}\mV^{\top}), \ldots, \pi_k(\mU_{i,:}\mV^{\top})
\]

\subsubsection{Personalized Visual-Configuration Recommendation}
The personalized ranking of the visual-configurations for user $i$ is inferred by $\mU_{i,:}\mZ^{\top}$ where $\mU_{i,:}$ is the embedding of user $i$ and $\mZ$ is the matrix of visual-configuration embeddings.
Hence, $\mU_{i,:}\mZ^{\top} \in \RR^{h}$ is an $h$-dimensional vector of weights indicating the likelihood/importance of each visual-configuration for that specific user $i$.
Let $\pi_1(\mU_{i,:}\mZ^{\top})$ denote the largest visual-configuration weight for user $i$.
Therefore, the top-$k$ visual-configuration weights for user $i$ is denoted as:
\[
\pi_1(\mU_{i,:}\mZ^{\top}), \pi_2(\mU_{i,:}\mZ^{\top}), \ldots, \pi_k(\mU_{i,:}\mZ^{\top})
\]

\subsubsection{Personalized Visualization Recommendation}
We now focus on the most complex and challenging problem of recommending complete visualizations personalized for a specific user $i \in [n]$.
A recommended visualization for user $i \in [n]$ consists of both the subset of \emph{attributes} $\mX^{(k)}$ from some dataset $\mX$ and the \emph{design choices} $C_{t}$ (a visual-configuration) for those attributes.
Given user $i$ along with an arbitrary visualization $\mathcal{V}=(\mX^{(k)},C_t)$ generated from some dataset $\mX$ of interest to user $i$, we derive a personalized user-specific score for visualization $\mathcal{V}$ (for user $i$) as,
\begin{align} \label{eq:personalized-vis-score-two-or-more-attributes}
\hat{y}(\mathcal{V}) = \mU_{i,:}\mZ_{t,:}^{\top} \prod_{\vx_j \in \mX^{(k)}} \mU_{i,:}\mV_{j,:}^{\top} 
\end{align}\noindent
where $\mX^{(k)}$ is the subset of attributes from the users dataset $\mX$ (hence, $|\mX^{(k)}|\leq |\mX|$) used in the visualization $\mathcal{V}$ 
and $C_{t} \in \mathbcal{C}$ is the visual-configuration of the visualization $\mathcal{V}$ being scored for user $i$.
Using Eq.~\ref{eq:personalized-vis-score-two-or-more-attributes}, we can predict the personalized visualization score 
$\hat{y}(\mathcal{V})$ for any arbitrary visualization $\mathcal{V}$ (for any dataset) and user $i \in [n]$.
For evaluation in Section~\ref{sec:exp-personalized-vis-rec}, we use Eq.~\ref{eq:personalized-vis-score-two-or-more-attributes} to score relevant and non-relevant visualizations for a specific user and dataset of interest.

\section{Deep Personalized Visualization Recommendation Models}\label{sec:neural-pvisrec}
We now introduce a deep neural network architecture for personalized visualization recommendation.
For this, we combine the previously proposed model with a deep multilayer neural network component to learn non-linear functions that capture complex dependencies and patterns between users and their visualization preferences.

\subsubsection{Neural PVisRec}
Given an arbitrary user $i$ and a visualization $\mathcal{V}=(\mX^{(k)}_{ij}, C_{t})$ to score from some new dataset of interest to that user, we first must decide on the input representation.
In this work, we leverage the user personalized embeddings learned in Section~\ref{sec:models} by concatenating the embedding of user $i$, visual configuration $t$, along with the embeddings for each attribute used in the visualization.
More formally, 
\begin{align}
\phi(\mathcal{V}\!=\!\langle\mX^{(k)}_{ij},C_t\rangle) = 
\begin{bmatrix}
\vu_i \\
\vz_t \\
\vv_{r_1} \\ \vdots \\ \vv_{r_s} 
\end{bmatrix}
\end{align}
where $\vu_i$ is the embedding of user $i$, 
$\vz_t$ is the embedding of the visual-configuration $C_{t}$, and
$\vv_{r_1}, \ldots, \vv_{r_s}$ are the embeddings of the attributes used in the visualization being scored for user $i$.
This can be written as,
\begin{align} \label{eq:PVisRec-MLP-input}
\phi(\mathcal{V}\!=\!\langle\mX^{(k)}_{ij},C_t\rangle) = 
\big[\,
\mU^{\top}\ve_i\;\,\,
\mZ^{\top} \ve_t\;\,\,
\mV^{\top} \ve_{r_1}\, \cdots\, \mV^{\top}\ve_{r_s}\, \big]^{\top}
\end{align}
where $\ve_i \in \RR^n$ (user $i$), 
$\ve_t \in \RR^{h}$ (visual-configuration $C_{t}$), and
$\ve_{r_1} \in \RR^{m}$ (attribute $r_1$) are the one-hot encodings of the user $i$, visual-configuration $t$, and attributes $r_1,...,r_s$ used in the visualization.
Note that $\mU \in \RR^{n \times d}$, $\mV \in \RR^{m \times d}$, $\mZ \in \RR^{h \times d}$, $\mY \in \RR^{k \times d}$.

The first \emph{neural personalized visualization recommendation} architecture that we introduce called \emph{Neural PVisRec} leverages the user, visual-configuration, and attribute embeddings from the PVisRec model in Section~\ref{sec:models} as input into a deep multilayer neural network with $L$ fully-connected layers,
\begin{align} 
\label{eq:MLP-input-layer}
\phi(\mathcal{V}\!=\!\langle\mX^{(k)}_{ij},C_t\rangle) &= \big[\,
\mU^{\top}\ve_i\;\,\,
\mZ^{\top} \ve_t\;\,\,
\mV^{\top} \ve_{r_1}\, \cdots\, \mV^{\top}\ve_{r_s}\, \big]^{\top}\\
\vq_1 &= \sigma_1(\mW_{1} \phi(\mathcal{V}) + \vb_{1}) \\ 
\vq_2 &= \sigma_2(\mW_{2} \vq_1 + \vb_{2}) \\ 
\vspace{-2mm}
&\;\,\vdots \nonumber \\ 
\vspace{-2mm}
\vq_L &= \sigma_L(\mW_{L} \vq_{L-1} + \vb_{L}) \\
\hat{y}_{\rm } &= \sigma(\vh^{\top} \vq_{L}) 
\label{eq:MLP-output-layer}
\end{align}
where $\mW_{L}$, $\vb_{L}$, and $\sigma_{L}$ are the weight matrix, bias vector, and activation function for layer $L$.
Further, $\hat{y}_{\rm } = \sigma(\vh^{\top} \vq_{L})$ (Eq.~\ref{eq:MLP-output-layer}) is the output layer where $\sigma$ is the output activation function and $\vh^{\top}$ denotes the edge weights of the output function.
For the hidden layers, we used ReLU as the activation function.
Note that if the visualization does not use all $s$ attributes, then we can pad the remaining unused attributes with zeros.
This enables the multi-layer neural network architecture to be flexible for visualizations with any number of attributes.
Eq.~\ref{eq:MLP-input-layer}-\ref{eq:MLP-output-layer} can be written more succinctly as
\begin{align} 
\label{eq:neural-PVisRec}
\hat{y}_{\rm } \,=\,
\sigma\big(\vh^{\top} \sigma_{L}(\mW_{L}(... \sigma_{1}(\mW_{1} 
\big[
\mU^{\top}\ve_i\;\,\,
\mZ^{\top} \ve_j\;\,\,
\mV^{\top} \ve_{r_1} \cdots\, \mV^{\top}\ve_{r_s} \big]^{\top} 
+ \vb_{1}) ...) + \vb_{L})\big)
\end{align}\noindent
where $\hat{y}_{\rm }$ is the predicted visualization score for user $i$.

\smallskip
\subsubsection{Neural PVisRec-CMF}
We also investigated a second neural approach for the personalized visualization recommendation problem.
This approach combines scores from PVisRec and Eq.~\ref{eq:neural-PVisRec}.
More formally, given user $i$ along with an arbitrary visualization $\mathcal{V}=(\mX^{(k)}_{ij},C_t)$ generated from some dataset $\mX_{ij}$ of interest to user $i$, we derive a personalized user-specific score for visualization $\mathcal{V}$ (for user $i$) as
$\hat{y}_{\rm PVisRec} = \mU_{i,:}\mZ_{t,:}^{\top} \prod_{\vx_j \in \mX^{(k)}} \mU_{i,:}\mV_{j,:}^{\top}$
where $\mX^{(k)}_{ij}$ is a subset of attributes used in the visualization $\mathcal{V}$ from the users dataset $\mX_{ij}$ (hence, $|\mX^{(k)}|\leq |\mX_{ij}|$) and $C_{t} \in \mathbcal{C}$ is the visual-configuration for visualization $\mathcal{V}$.
Then, we have
\begin{align}\label{eq:neural-PVisRec-CMF}
\hat{y} \,=\, 
(1-\alpha)\Bigg(\mU_{i}\mZ_{t}^{\top} \prod_{\vx_j \in \mX^{(k)}} \mU_{i}\mV_{j}^{\top}\Bigg) \;+\; 
\alpha \hat{y}_{\rm dnn}
\end{align}
where 
$\hat{y}_{\rm dnn} = \sigma\big(\vh^{\top} \sigma_{L}(\mW_{L}(... \sigma_{1}(\mW_{1} 
\phi(\mathcal{V})
+ \vb_{1}) ...) + \vb_{L})\big)$
with 
$\phi(\mathcal{V})=
\big[
\mU^{\top}\ve_i\;\,
\mZ^{\top} \ve_t\;\,
\mV^{\top} \ve_{r_1} \cdots\, \mV^{\top}\ve_{r_s} \big]^{\top}
$
and
$\alpha \in (0,1)$ is a hyperparameter that controls the influence of the models on the final predicted score of the visualization for user $i$.

All layers of the various neural architectures for our personalized visualization recommendation problem use ReLU nonlinear activation.
Unless otherwise mentioned, we used three hidden layers and optimized model parameters using mini-batch Adam with a learning rate of 0.001.
We designed the neural network structure such that the bottom layers are the widest and each successive layer has 1/2 the number of neurons.
For fairness, the last hidden layer is set to the embedding size.
Hence, if the embedding size is 8, then the architecture of the layers is $32 \rightarrow 16 \rightarrow 8$.

\subsubsection{Training}
The user-centric visualization training corpus $\mathbcal{D} = \{\mathbcal{X}_i,\mathbb{V}_i\}_{i=1}^{n}$ for \emph{personalized} visualization recommendation consists of user-level training data for $n$ users where for each user $i \in [n]$ we have a set of datasets $\mathbcal{X}_{i}=\{\mX_{i1},\ldots,\mX_{ij},\ldots\}$ of interest to that 
user along with user $i$'s ``relevant'' (generated, liked, clicked-on) visualizations $\mathbb{V}_{i}=\{\mathbcal{V}_{i1},\ldots,\mathbcal{V}_{ij},\ldots\}$ for each of those datasets.
For each user $i \in [n]$ and dataset $\mX_{ij} \in \mathbcal{X}_{i}$ of interest to user $i$, there is a set $\mathbcal{V}_{ij}=\{\ldots,\mathcal{V}=(\mX_{ij}^{(k)}\!, \mathcal{C}_{ijk}),\ldots\}$ of relevant (positive) visualizations for that user, and we also leverage a sampled set of non-relevant (negative) visualizations $\mathbcal{V}_{ij}^{-}$ for that user $i$ and dataset $\mX_{ij} \in \mathbcal{X}_{i}$.
Therefore, the set of training visualizations for user $i \in [n]$ and dataset $\mX_{ij} \in \mathbcal{X}_{i}$ is $\mathbcal{V}_{ij} \cup \mathbcal{V}_{ij}^{-}$ and $Y_{ijk} \in \{0,1\}$ denotes the ground-truth label of visualization $\mathcal{V}=(\mX_{ij}^{(k)}\!, \mathcal{C}_{ijk}) \in \mathbcal{V}_{ij} \cup \mathbcal{V}_{ij}^{-}$.
Hence, $Y_{ijk}=1$ indicates a user-relevant (positive) visualization for user $i$ whereas $Y_{ijk}=0$ indicates a non-relevant visualization for that user, \ie, $\mathcal{V}=(\mX_{ij}^{(k)}\!, \mathcal{C}_{ijk}) \in \mathbcal{V}_{ij}^{-}$.
The goal is to have the model score $\hat{Y}_{ijk} \in [0,1]$ each training visualization $\mathcal{V}=(\mX_{ij}^{(k)}\!, \mathcal{C}_{ijk}) \in \mathbcal{V}_{ij} \cup \mathbcal{V}_{ij}^{-}$ for a user $i$ as close as possible to the ground-truth label $Y_{ijk}$.
The neural personalized visualization recommendation model is learned by optimizing the likelihood of model scores for all visualizations of each user.
Given a user $i \in [n]$ and the model parameters $\Theta$, the likelihood is
\begin{equation}\label{eq:user-level-likelihood}
\mathrm{P}(\hat{\mathbb{V}}_i^{-}, \mathbb{V}_i| \Theta) \,= 
\prod_{j=1}^{|\mathbcal{X}_i|} \;\,
\prod_{(\mX_{ij}^{(k)}\!, \mathcal{C}_{ijk})\in\mathbcal{V}_{ij}} \!\hat{Y}_{ijk} 
\prod_{(\mX_{ij}^{(k)}\!,\mathcal{C}_{ijk})\in \hat{\mathbcal{V}}_{ij}^{-}} \!\!\Big(1 - \hat{Y}_{ijk}\Big), 
\;\;
\text{for } i=1,\ldots,n
\end{equation}
where $\hat{Y}_{ijk}$ is the predicted score of a visualization $\mathcal{V} = (\mX_{ij}^{(k)}, \mathcal{C}_{ijk})$ for 
user $i$ and dataset $j$ ($\mX_{ij} \in \mathbcal{X}_i$).
Naturally, the goal is to obtain $\hat{Y}_{\mathcal{V}}$ such that it is as close as possible to the actual ground-truth $Y_{\mathcal{V}}$.
Taking the negative log of the likelihood in Eq.~\ref{eq:user-level-likelihood} and summing over all $n$ users and their sets of relevant visualizations $\mathbcal{V}_{ij}$ from $|\mathbcal{X}_i|$ different datasets give us the total loss $\mathbb{L}$. 
\begin{equation} \label{eq:objective-neural-PVisRec-negative-log-LL}
\begin{aligned}
\mathbb{L} &= \sum_{i=1}^{n}
\sum_{j=1}^{|\mathbcal{X}_i|} 
\Bigg(-\sum_{(\mX_{ij}^{(k)}\!, \mathcal{C}_{ijk})\in\mathbcal{V}_{ij}}\log \hat{Y}_{ijk} - 
\sum_{(\mX_{ij}^{(k)}\!,\mathcal{C}_{ijk})\in \hat{\mathbcal{V}}_{ij}^{-}}\log (1 - \hat{Y}_{ijk})\Bigg) \\
&= - \sum_{i=1}^{n}
\sum_{j=1}^{|\mathbcal{X}_i|} 
\sum_{(\mX_{ij}^{(k)}\!,\mathcal{C}_{ijk})\in \mathbcal{V}_{ij} \cup \hat{\mathbcal{V}}_{ij}^{-}} Y_{ijk}\log \hat{Y}_{ijk} + (1 - Y_{ijk}) \log (1 - \hat{Y}_{ijk})
\end{aligned}
\end{equation}\noindent
where the objective function above is minimized via stochastic gradient descent (SGD) to update the model parameters $\Theta$ in $\mathcal{M}$.

\section{Benchmark Data for Personalized Visualization Recommendation}
\label{sec:dataset}
Since this is the first work that addresses the \emph{personalized} visualization recommendation problem, there were not any existing public datasets that could be used directly for our problem.
Recent works have ignored the user information~\cite{vizml,ML-based-Vis-Rec} that details the ``author'' of the visualization, which is required in this work for user-level personalization.
As an aside, VizML~\cite{vizml} discarded all user information and only kept the attributes used in an actual visualization (and therefore did not consider datasets as well).
In this work, since we focus on the personalized visualization recommendation problem, we derive a user-centered dataset where for each user we know their datasets, visualizations, attributes, and visualization-configurations used.
We started from the raw Plot.ly community feed data.\footnote{http://vizml-repository.s3.amazonaws.com/plotly\_full.tar.gz}
For the personalized visualization recommendation problem, we first extract the set of all $n$ users in the visualization corpus.
For each user $i \in [n]$, we then extract the set of datasets $\mathbcal{X}_i$ of interest to that user. 
These are the datasets that user $i$ has generated at least one visualization. 
Depending on the visualization corpus data, this could also be other types of 
user feedback such as a visualization that a user liked or clicked.
Next, we extract the set of user-preferred visualizations $\mathbcal{V}_{ij}$ for each of the datasets $\mX_{ij} \in \mathbcal{X}_i$ of interest to user $i$.
Hence, $\mathbcal{V}_{ij}$ is the set of visualizations generated (or liked, clicked, ...) by user $i$ for dataset $j$ ($\mX_{ij}$).
Every visualization $\mathcal{V} \in \mathbcal{V}_{ij}$ preferred by user $i$ also obviously contains the attributes from dataset $\mX_{ij} \in \mathbcal{X}_i$ used in the visualization (\ie, the attributes that map to the x, y, binning, color, and so on).

In Table~\ref{table:dataset-statistics}, we report statistics about the personalized visualization corpus used in our work, including the number of users, attributes, datasets, visualizations, and visualization-configurations extracted from all the user-generated visualizations, and so on.
The corpus $\mathbcal{D}=\{\mathbcal{X}_i, \mathbb{V}_i\}_{i=1}^{n}$ for learning individual personalized visualization recommendation models consists of a total of $n=17,469$ users with $|\!\bigcup_{i=1}^n \mathbcal{X}_i|=94,419$ datasets used by those users.
Further, there are $m=2,303,033$ attributes among the $94,419$ datasets of interest by the $17.4k$ users.
Our user-centric visualization training corpus $\mathbcal{D}$ has a total of $|\!\bigcup_{i=1}^n \mathbb{V}_i| = $ 32,318 relevant visualizations generated by the $n=17.4k$ users with an average of 1.85 relevant visualizations per user.
Each user in the corpus has an average of 5.41 datasets 
and each dataset has an average of 24.39 attributes.
From the 32.3k user-relevant visualizations from the $17.4k$ users, we extracted a total of $|\mathbcal{C}|=686$ unique visual-configurations. 
To further advance research on personalized visualization recommender systems, we have made the user-level plot.ly data that we used for studying the personalized visualization recommendation problem (introduced in Section~\ref{sec:problem-recommending-personalized-vis}) 
publicly accessible at: 

\begin{center}
\href{http://networkrepository.com/personalized-vis-rec}{http://networkrepository.com/personalized-vis-rec}
\end{center}

\noindent
We have also made the graph representations used in our personalized visualization recommendation framework publicly accessible at  \href{http://networkrepository.com/personalized-vis-rec-graphs}{http://networkrepository.com/personalized-vis-rec-graphs}

\begin{table}[t!]
\caption{Personalized visualization recommendation data corpus. These user-centric dataset is used for learning personalized visualization recommendation models for individual users.}
\label{table:dataset-statistics}
\vspace{-2mm}
\small
\setlength{\tabcolsep}{10.3pt} 
\def\arraystretch{1.05} 
\begin{tabular}{l r  }
\toprule\textsc{\# Users} & 17,469 \\
\textsc{\# Datasets} & 94,419 \\
\textsc{\# Attributes} & 2,303,033 \\
\textsc{\# Visualizations} & 32,318 \\
\textsc{\# Vis. Configs} & 686 \\
\textsc{\# Meta-features} & 1006 \\
\midrule
\textsc{mean \# attr. per dataset} & 24.39 \\
\textsc{mean \# attr. per user} & 51.63 \\
\textsc{mean \# vis. per user} & 1.85 \\
\textsc{mean \# datasets per user} & 5.41 \\
\midrule
\textsc{Density ($\mA$)} & <0.0001 \\
\textsc{Density ($\mC$)} & <0.0001 \\
\textsc{Density ($\mD$)} & <0.0001 \\
\textsc{Density ($\mM$)} & 0.4130 \\
\bottomrule
\end{tabular}
\end{table}

\section{Experiments} \label{sec:exp}
To investigate the effectiveness of the \emph{personalized visualization recommendation} approach, we design experiments to answer the following research questions:
\begin{itemize}[leftmargin=*]

\item \textbf{RQ1:} 
Given a user and a new dataset of interest to that user, 
can we accurately recommend the top most relevant visualizations for that specific user (Section~\ref{sec:exp-personalized-vis-rec})?

\item \textbf{RQ2:} 
How does our user-level personalized visualization recommendations compare to the non-personalized global recommendations (Section~\ref{sec:exp-personalized-vs-nonpersonalized-vis-rec})?

\item \textbf{RQ3:}
Can we significantly reduce the space requirements of our approach by trading off a small amount of accuracy for a large improvement in space (Section~\ref{sec:exp-space-efficient-results-meta-feature-embeddings})?

\item \textbf{RQ4:} 
Does the neural personalized visualization recommendation models further improve the performance when incorporating a 
multilayer deep neural network component (Section~\ref{sec:exp-adding-DL-component})?

\end{itemize}

\begin{table}[h!]
\caption{Personalized Visualization Recommendation Results.
Note $d=10$.
See text for discussion.
}
\label{table:vis-rec-100k}
\vspace{-2mm}
\footnotesize
\begin{tabular}{l ccccc   ccccc   }
\toprule
 &\multicolumn{5}{c}{HR@K} &\multicolumn{5}{c}{NDCG@K} \\
\cmidrule(lr){2-6} 
\cmidrule(lr){7-11} 
 \textbf{Model} & @1 & @2 & @3 & @4 & @5 & @1 & @2 & @3 & @4 & @5 \\
\midrule
 VizRec & N/A & N/A & N/A & N/A & N/A & N/A & N/A & N/A & N/A & N/A \\
  VisPop & 0.186 & 0.235 & 0.255 & 0.271 & 0.289 & 0.181 & 0.214 & 0.224 & 0.231 & 0.238 \\
  VisConfigKNN & 0.026 & 0.030 & 0.038 & 0.055 & 0.089 & 0.016 & 0.021 & 0.026 & 0.034 & 0.048 \\
  VisKNN & 0.147 & 0.230 & 0.297 & 0.372 & 0.449 & 0.143 & 0.195 & 0.227 & 0.257 & 0.286 \\
  eALS & 0.304 & 0.395 & 0.426 & 0.441 & 0.449 & 0.302 & 0.360 & 0.376 & 0.382 & 0.385 \\
  MLP & 0.218 & 0.452 & 0.601 & 0.671 & 0.715 & 0.211 & 0.357 & 0.435 & 0.465 & 0.483 \\
  PVisRec & \textbf{0.630} & \textbf{0.815} & \textbf{0.876} & \textbf{0.906} & \textbf{0.928} & \textbf{0.624} & \textbf{0.743} & \textbf{0.775} & \textbf{0.788} & \textbf{0.796} \\
\bottomrule
\end{tabular}
\end{table}

\subsection{Personalized Visualization Recommendation Results}\label{sec:exp-personalized-vis-rec}

\subsubsection{Experimental setup}
Now we evaluate the system for recommending personalized visualizations to a user.
Given an arbitrary user, we know the visualization(s) they preferred for each of the datasets of interest to them.
Therefore, we can quantitatively evaluate the proposed approach for personalized visualization recommendation.
For each user, we randomly select one of their datasets where the user has manually created at least two visualizations (treated as positive examples), and randomly select one of those positive visualizations to use for testing, and the other positive instances are used for training and validation.
This is similar to leave-one-out evaluation which is widely used in traditional user-item recommender systems~\cite{eALS}.
However, in our case, we have thousands of datasets, and for each dataset there are a large and completely \emph{disjoint} set of possible visualizations to recommend to that user.\footnote{The set of candidate visualizations for a specific dataset are not only disjoint (\ie, completely different from any other set of visualizations generated from another dataset), but the amount of possible visualizations for a given dataset are exponential in the number of attributes, possible design choices, and so on, making this problem unique and fundamentally challenging.} 
Since it is too computationally expensive to rank all visualizations for every user (and every dataset of interest) during evaluation, we randomly sample 19 visualizations that were not created by the users.
This gives us a total of 20 visualizations per user (1 relevant + 19 non-relevant visualizations) to use for evaluation of the personalized visualization recommendations from our proposed models.
Using this held-out set of user visualizations, we evaluate the ability of the proposed approach to recommend these held-out relevant visualizations to the user (which are visualizations the user actually created), among the exponential amount of alternative visualizations (that arise for a single dataset of interest) from a set of attributes and sets of design choices (\eg, chart-types, ...).
In particular, given a user $i$ and a dataset of interest to that user, we use the proposed approach to recommend the top-$k$ visualizations personalized for that specific user and dataset.
To quantitatively evaluate the personalized ranking of visualizations given by the proposed personalized visualization recommendation models, we use rank-based evaluation metrics including Hit Ratio at $K$ (HR@K) and Normalized Discounted Cumulative Gain (NDCG@K)~\cite{eALS}.
Intuitively, HR@K quantifies whether the held-out relevant (user generated) visualization appears in the top-$K$ ranked visualizations or not.
Similarly, NDCG@K takes into account the position of the relevant (user generated) visualization in the top-$K$ ranked list of visualizations, by assigning larger scores to visualizations ranked more highly in the list.
For both HR@K and NDCG@K, we report $K=1,\ldots,5$ unless otherwise mentioned.

Therefore, given a user $i$ along with the set of relevant and non-relevant visualizations $\mathbcal{V}_{ij} \cup \mathbcal{V}_{ij}^{-}$ for that user and their dataset $\mX_{ij} \in \mathbcal{X}_i$ of interest, we derive a score for each of the visualizations $\mathcal{V} \in (\mathbcal{V}_{ij} \cup \mathbcal{V}_{ij}^{-})$ where $|\mathbcal{V}_{ij}| + |\mathbcal{V}_{ij}^{-}|=20$.
An effective personalized visualization recommender will assign a larger score to the relevant visualizations and smaller scores to the non-relevant visualizations, hence, the relevant visualizations will show up first, followed by the non-relevant visualizations (which should appear further down the list).
Unless otherwise mentioned, we use $d=10$ as the embedding size and use the full meta-feature matrix $\mM$.
For the neural variants of our approach, we use $\alpha=0.5$.

\begin{table}[h!]
\caption{Ablation study results for different variants of our personalized visualization recommendation approach.}
\label{table:vis-rec-variants-ablation-10k}
\vspace{-2mm}
\footnotesize
\begin{tabular}{l ccccc   ccccc   }
\toprule
 &\multicolumn{5}{c}{HR@K} &\multicolumn{5}{c}{NDCG@K} \\ 
\cmidrule(lr){2-6} 
\cmidrule(lr){7-11} 
 \textbf{Model} & @1 & @2 & @3 & @4 & @5 & @1 & @2 & @3 & @4 & @5 \\
\midrule
  PVisRec ($\mA$,$\mC$,$\mM$ only) & 0.307 & 0.416 & 0.470 & 0.488 & 0.501 & 0.306 & 0.374 & 0.401 & 0.410 & 0.415 \\
  PVisRec ($\mA$,$\mC$,$\mD$ only) & 0.414 & 0.474 & 0.537 & 0.610 & 0.697 & 0.384 & 0.435 & 0.450 & 0.457 & 0.460 \\
  PVisRec & \textbf{0.630} & \textbf{0.815} & \textbf{0.876} & \textbf{0.906} & \textbf{0.928} & \textbf{0.624} & \textbf{0.743} & \textbf{0.775} & \textbf{0.788} & \textbf{0.796} \\
\bottomrule
\end{tabular}
\end{table}

\subsubsection{Baselines}
Since the personalized visualization recommendation problem introduced in Section~\ref{sec:problem} is new, 
there are not any existing vis. rec. methods that can be directly applied to solve it.
For instance, VizRec~\cite{vizrec} is the closest existing approach, though is unable to be used 
since it explicitly assumes a single dataset where users provide feedback about visualizations pertaining to that dataset of interest.
However, in our problem formulation and corpus $\mathbcal{D}=\{(\mathbcal{X}_i, \mathbb{V}_i)\}_{i=1}^n$, every user $i \in [n]$ can have their own set of datasets $\mathbcal{X}_i$ that are not shared by any other user.
In such cases, it is impossible to use VizRec.
Nevertheless, we adapted a wide variety of methods to use as baselines for evaluation.
We now briefly summarize these methods below:

\begin{itemize}
\item \textbf{VisPop}: Given a visualization $\mathcal{V}$ with attributes $\mX^{(k)}$ and visual-configuration $\mathcal{C} \in \mathbcal{C}$, the score of visualization $\mathcal{V}$ is $\phi(V) = f(\mathcal{C}) \prod_{\vx \in \mX^{(k)}} f(\vx)$
where $f(\vx)$ is the frequency of attribute $\vx$ 
(sum of the columns of $\mA$) 
and $f(\mathcal{C})$ is the frequency of visual-configuration $\mathcal{C}$.
Hence, the score given by VisPop is a product of the frequencies of the underlying visualization components, \ie, visual-configuration and attributes used in the visualization being scored.

\item \textbf{VisKNN}: 
This is the standard item-based collaborative filtering method adapted for the visualization recommendation problem.
Given a visualization $\mathcal{V}$ with attributes $\mX^{(k)}$ and visual-configuration $\mathcal{C} \in \mathbcal{C}$, then we score $\mathcal{V}$ by taking the mean score of the visual configurations most similar to $\mathcal{C}$, 
along with the mean score of the top attributes most similar to each of the attributes used in the visualization.

\item \textbf{VisConfigKNN}: 
This approach is similar to VisKNN, but uses only the visual-configuration matrix to score the visualizations.

\item \textbf{eALS}: This is an adapted version of the state-of-the-art MF method used for item recommendation in~\cite{eALS}. We adapted it for our visualization recommendation problem by minimizing squared loss while treating all unobserved user iterations between attributes and visual-configurations as negative examples, which are weighted non-uniformly by the frequency of attributes and visual-configurations.

\item \textbf{MLP}:
We used three hidden layers and optimized model parameters using mini-batch Adam with a learning rate of 0.001.
For the activation functions of the MLP layers, we used ReLU.
For fairness, the last hidden layer is set to the embedding size.

\item \textbf{VizRec}~\cite{vizrec}:
For each dataset, this approach constructs a user-by-visualization matrix and uses it to obtain the average overall rating among the similar users of a visualization, where a user is similar if it has rated a visualization preferred by the active user. 
VizRec assumes a single dataset and is only applicable when there are a large number of users that have rated visualizations from the \emph{same} dataset.
\end{itemize}

\begin{figure*}[t!]
\centering
\subfigure{\includegraphics[width=0.48\linewidth]{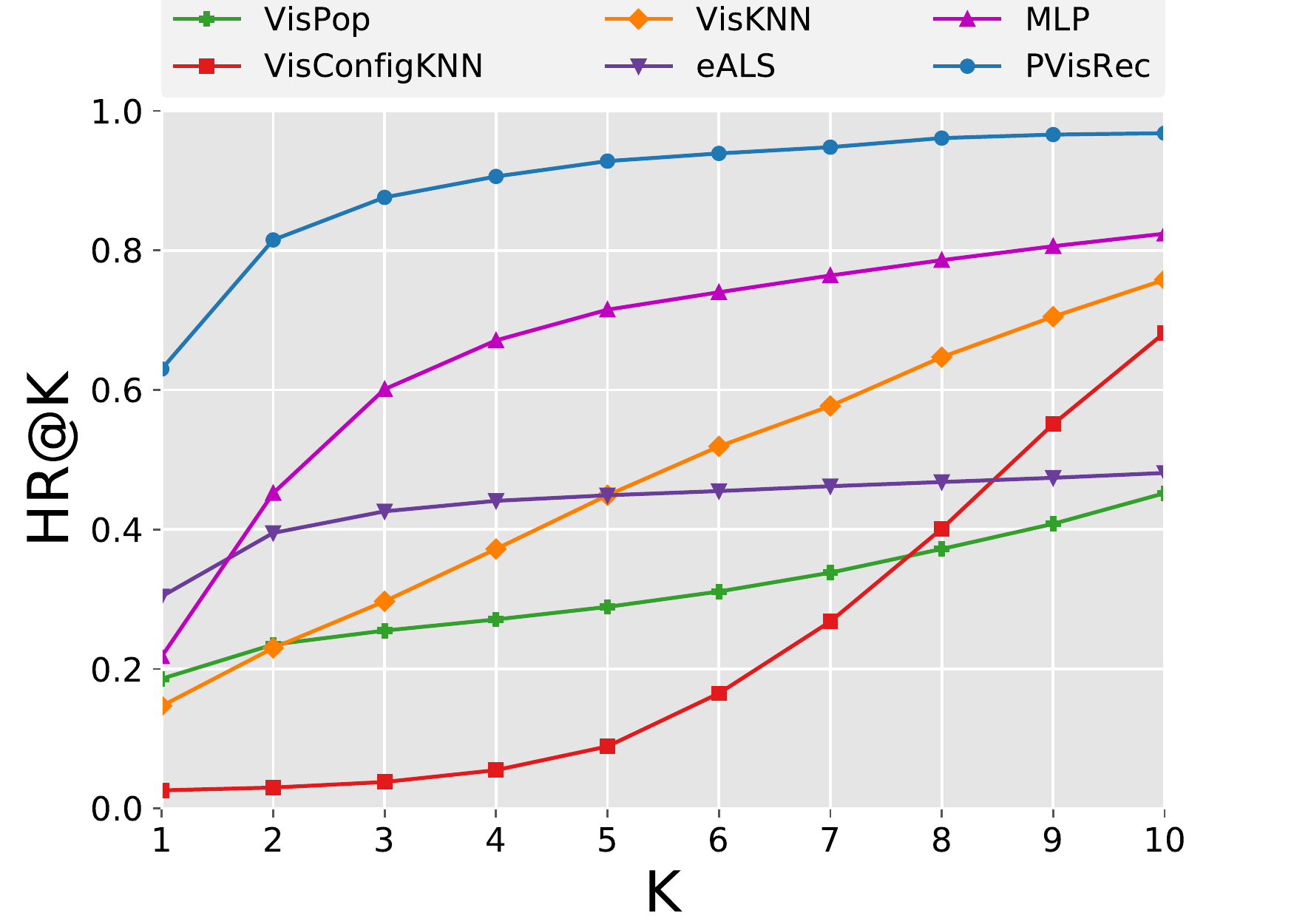}}
\subfigure{\includegraphics[width=0.48\linewidth]{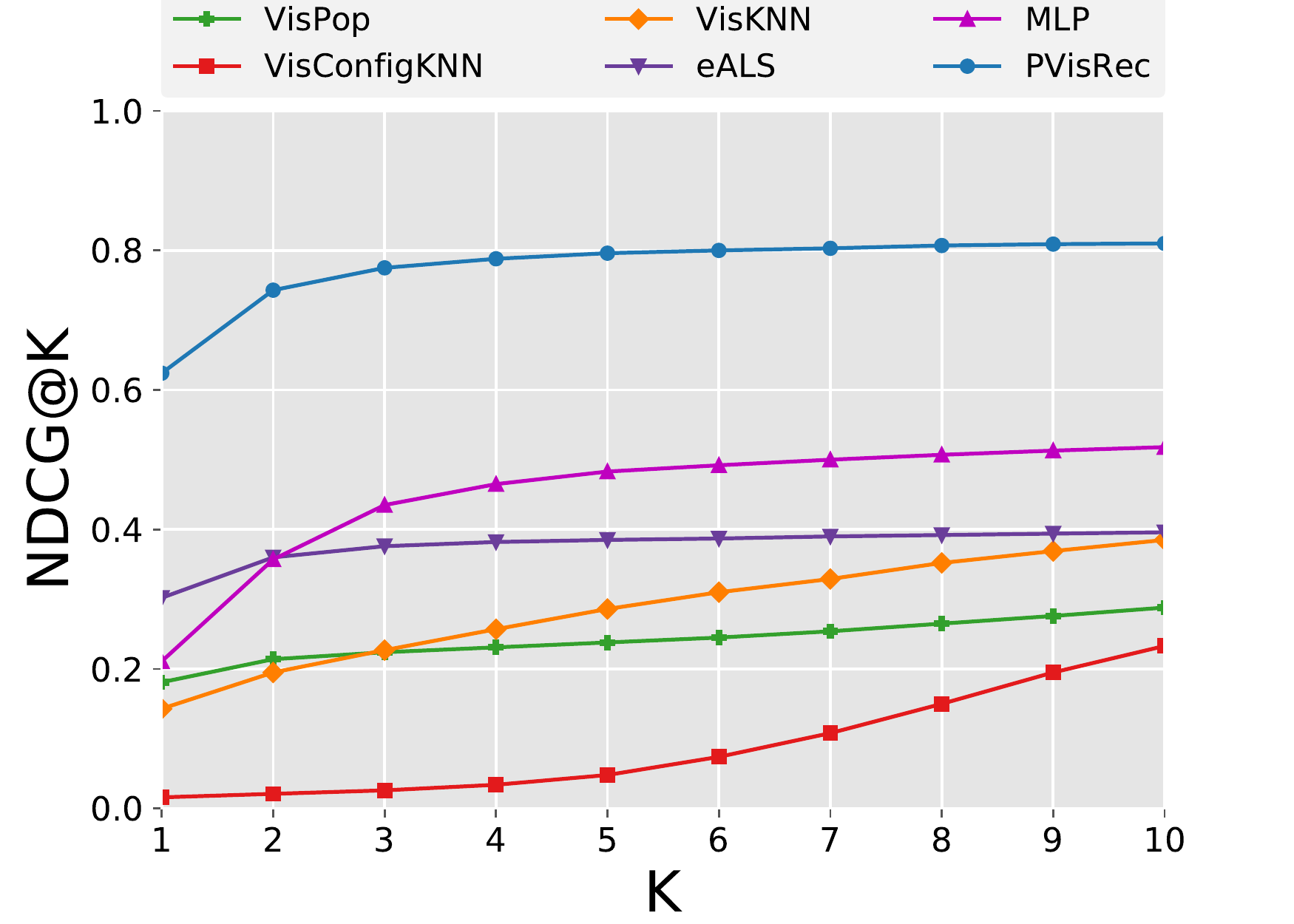}}
\vspace{-1mm}
\caption{Evaluation of top-K personalized visualization recommendations.}
\label{fig:results-hit-and-ndcg}
\end{figure*}

\begin{figure*}[t!]
\centering
\includegraphics[width=0.90\linewidth]{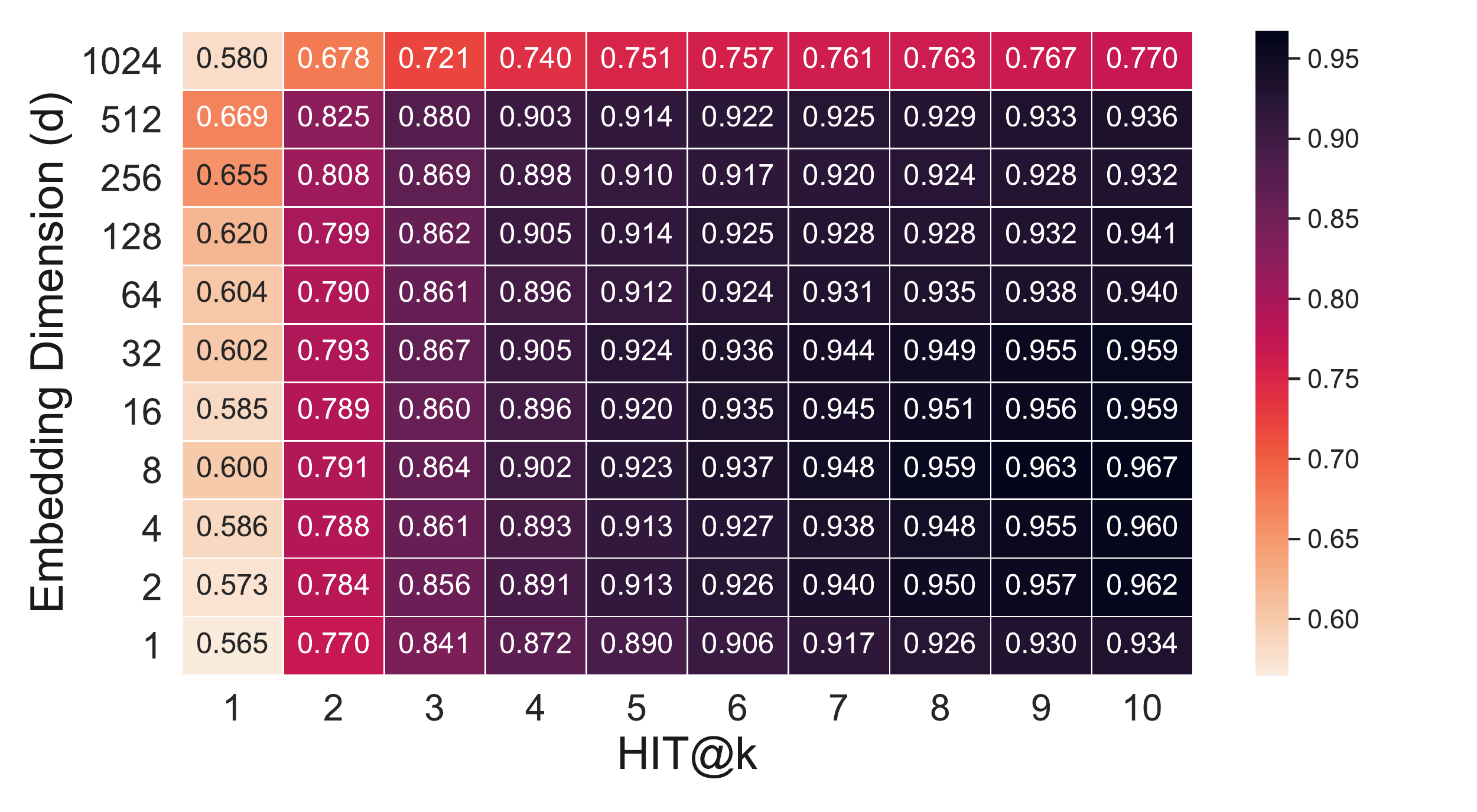}
\vspace{-3mm}
\caption{Ablation study results for personalized visualization recommendation with varying embedding dimensions $d \in \{2^0,\ldots,2^{10}\}$ and HIT@k for $k=1,\ldots,10$. See text for discussion.}
\label{fig:vary-embedding-dim-hitk}
\end{figure*}

\subsubsection{Results}
We provide the results in Table~\ref{table:vis-rec-100k}.
Overall, the proposed approach, PVisRec, significantly outperforms the baseline methods by a large margin as shown in Table~\ref{table:vis-rec-100k}.
Strikingly, PVisRec consistently achieves the best HR@K and NDCG@K across all $K=1,2,\ldots,5$.
From Table~\ref{table:vis-rec-100k}, we see that PVisRec achieves a mean relative improvement of $107.2\%$ and $106.6\%$ over the best performing baseline method (eALS) for HIT@1 and NDCG@1, respectively.
Comparing HIT@5 and NDCG@5, PVisRec achieves a mean improvement of $29.8\%$ and $64.9\%$ over the next best performing method (MLP).
As an aside, VizRec is the only approach proposed for ranking visualizations.
All other methods used in our comparison are new and to the best of our knowledge have never been extended for ranking and recommending visualizations.
Recall that VizRec in itself solves a different problem, but we point out the above since it is clearly the closest.
As discussed in Section~\ref{sec:related-work}, all of the assumptions required by VizRec are unrealistic in practice.
This is also true when using VizRec for our problem and corpus $\mathbcal{D}=\{(\mathbcal{X}_i, \mathbb{V}_i)\}_{i=1}^n$ where every user $i \in [n]$ can have their own set of datasets $\mathbcal{X}_i$ that are not shared by any other user.
In such cases, we use ``N/A'' to denote this fact.
This is due to the VizRec assumption that there is a single dataset of interest by all $n$ users, and every user has given many different preferences on the relevant visualizations generated for that specific dataset.
All of these assumptions are violated in our problem.
Figure~\ref{fig:results-hit-and-ndcg} shows the mean performance of the top-$K$ visualization recommendations for $K=1,2,\ldots,10$.
These results demonstrate the effectiveness of our user personalized visualization recommendation approach as we are able to successfully recommend users the held-out visualizations that they previously created.

\subsubsection{Ablation Study Results}
Previously, we observed that PVisRec significantly outperforms other methods for the personalized visualization recommendation problem.
To understand the importance of the different model components of PVisRec, we investigate a few different variants of our personalized visualization recommendation model.
The first variant called PVisRec ($\mA, \mC, \mM$ only) does not use the attribute by visual-configuration graph represented by the sparse adjacency matrix $\mD$
whereas the second variant called PVisRec ($\mA, \mC, \mD$ only) does not use the dense meta-feature matrix $\mM$ for learning.
This is in contrast to PVisRec that uses $\mA, \mC, \mD$ and $\mM$.
In Table~\ref{table:vis-rec-variants-ablation-10k}, we see that both variants perform worse than PVisRec, indicating the importance of using all the graph representations
for learning the personalized visualization recommendation model.
Further, PVisRec ($\mA, \mC, \mD$ only) outperforms the other variant across both ranking metrics and across all $K$.
This suggests that $\mD$ may be more important for learning than $\mM$.
Nevertheless, the best personalized visualization recommendation performance is obtained when both $\mD$ and $\mM$ are used along with $\mA$ and $\mC$.
Finally, these two simpler variants still perform better than the baselines for HR@1 and NDCG@1 as shown in Table~\ref{table:vis-rec-100k}.

\begin{figure*}[t!]
\centering
\includegraphics[width=0.90\linewidth]{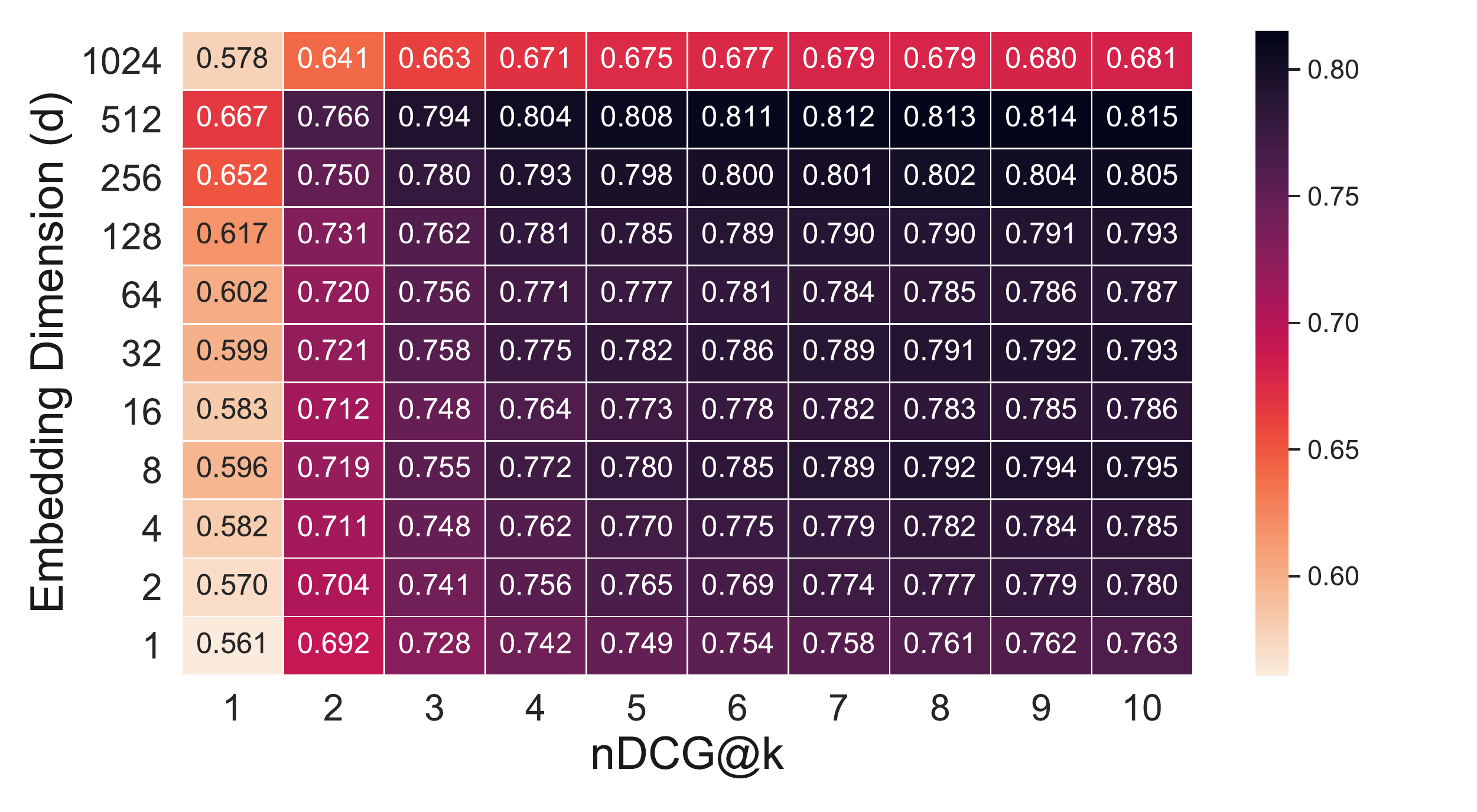}
\vspace{-3mm}
\caption{Ablation study results for personalized visualization recommendation with varying embedding dimensions $d \in \{2^0,\ldots,2^{10}\}$ and nDCG@k for $k=1,\ldots,10$. See text for discussion.}
\label{fig:vary-embedding-dim-ndcgk}
\end{figure*}

To understand the effect of the embedding size $d$ 
on the performance of our personalized visualization recommendation approach, we vary the dimensionality of the embeddings $d$ from 1 to 1024.
In these experiments, we use PVisRec with the full-rank meta-feature matrix $\mM$.
In Figure~\ref{fig:vary-embedding-dim-hitk}, we show results for the personalized visualization recommendation problem using our PVisRec approach with varying embedding dimensions (size) $d \in \{2^0,\ldots,2^{10}\}$ and HR@K for $k=1,\ldots,10$.
In addition, we also provide results in Figure-\ref{fig:vary-embedding-dim-ndcgk} for NDCG@K for $k=1,\ldots,10$ while varying the embedding size $d \in \{2^0,\ldots,2^{10}\}$.
This experiment uses the original meta-feature matrix $\mM$ and not the compressed meta-feature embedding (MFE) matrix.
For both HR@K and NDCG@K, we observe in Figure~\ref{fig:vary-embedding-dim-hitk}-\ref{fig:vary-embedding-dim-ndcgk} that performance typically increases as a function of the embedding dimension $d$.
We also observe that for HIT@1 and nDCG@1, the best performance is achieved when $d=512$, which is 0.669 and 0.667, respectively. 
This holds for all k for both HR@K and NDCG@K as shown in Figure~\ref{fig:vary-embedding-dim-hitk}-\ref{fig:vary-embedding-dim-ndcgk}.
Furthermore, when $d$ becomes too large, we observe a large drop in performance, which is due to overfitting.
For instance, in Figure~\ref{fig:vary-embedding-dim-hitk}, we see that when $d=1024$ we have HIT@1 of 0.580 compared to 0.669 for $d=512$.

\begin{table}[h!]
\caption{Results comparing Non-personalized vs. Personalized Visualization Recommendation.}
\label{table:vis-rec-10k-ML-vs-Personalized}
\footnotesize
\begin{tabular}{l ccccc  ccccc  }
\toprule
&\multicolumn{5}{c}{HR@K} &\multicolumn{5}{c}{NDCG@K} \\
\cmidrule(lr){2-6} 
\cmidrule(lr){7-11} 
\textbf{Model} & @1 & @2 & @3 & @4 & @5 & @1 & @2 & @3 & @4 & @5 \\
\midrule
Non-personalized & 0.151 & 0.248 & 0.319 & 0.373 & 0.404 & 0.145 & 0.209 & 0.244 & 0.268 & 0.280 \\
Personalized & \textbf{0.630} & \textbf{0.815} & \textbf{0.876} & \textbf{0.906} & \textbf{0.928} & \textbf{0.624} & \textbf{0.743} & \textbf{0.775} & \textbf{0.788} & \textbf{0.796} \\
\bottomrule
\end{tabular}
\end{table}

\subsection{Comparing Personalized vs. Non-personalized Visualization Recommendation}\label{sec:exp-personalized-vs-nonpersonalized-vis-rec}
To answer RQ2, we compare the personalized visualization recommendation model (PVisRec) to a non-personalized ML model.
More specifically, we compare the user-specific personalized visualization recommendation model (PVisRec) to a global non-personalized ML-based method that does not leverage a user-specific personalized model for each user.
For fairness, we simply leverage the specific user embedding for the personalized model, and for the non-personalized model we simply derive an aggregate global embedding of a typical user, and leverage this global non-personalized model to rank the visualizations.
More formally, the non-personalized ML-based approach uses a global user embedding derived as,
\begin{align} \label{eq:ml-based-global}
\vu_{g} = \frac{1}{n} \sum_{i=1}^n \mU_i
\end{align}\noindent
where $\vu_{g}$ is called the global user embedding and represents the centroid of the user embeddings from PVisRec.
Everything else remains the same as the personalized visualization recommendation approach.
More formally, given a user $i$ along with an arbitrary visualization $\mathcal{V}=(\mX^{(k)},C_t)$ generated from some dataset $\mX$, we derive a score for the visualization $\mathcal{V}$ using the global user embedding $\vu_{g}$ from Eq.~\ref{eq:ml-based-global} as follows:
\begin{align} \label{eq:non-personalized-vis-score-global-typical-user}
\phi_{g}(V) = \vu_{g}\mZ_{t,:}^{\top} \prod_{\vx_j \in \mX^{(k)}} \vu_{g}\mV_{j,:}^{\top} 
\end{align}\noindent
where $\mX^{(k)}$ is a subset of attributes used in the visualization $\mathcal{V}$ from the dataset $\mX$ (hence, $|\mX^{(k)}|\leq |\mX|$) and $\mathcal{C}_{t} \in \mathbcal{C}$ is the visual-configuration of $\mathcal{V}$.
Hence, instead of leveraging user $i$'s personalized visualization recommendation model to obtain a user personalized score for visualization $\mathcal{V}$ (that is $\phi(V) = \mU_{i,:}\mZ_{t,:}^{\top} \prod_{\vx_j \in \mX^{(k)}} \mU_{i,:}\mV_{j,:}^{\top}$), we replace $\mU_{i,:}$ with 
the global user embedding $\vu_g$ representing a ``typical'' user.
Results are provided in Table~\ref{table:vis-rec-10k-ML-vs-Personalized}.
For both models in Table~\ref{table:vis-rec-10k-ML-vs-Personalized}, we use the same experimental setup from Section~\ref{sec:exp-personalized-vis-rec}.
This PVisRec model is used for learning $\mU$, then Eq.~\ref{eq:ml-based-global} is used for the non-personalized model.
Notably, the non-personalized approach that uses the same global user model for all users performs significantly worse (as shown in Table~\ref{table:vis-rec-10k-ML-vs-Personalized}) compared to the user-level personalized approach that leverages the appropriate learned model to personalize the ranking of visualizations with respect to the user at hand.
This demonstrates the significance of learning individual models for each user that are personalized based on the users attribute/data preferences along with their visual design choice preferences \textbf{(RQ2)}.

\begin{table}[h!]
\caption{Space vs. Accuracy Trade-off Results using Meta-Feature Embeddings (MFE).
Results for the space-efficient variants of our personalized visualization recommendation methods that use meta-feature embeddings.
In particular, we set $d=10$ and vary the dimensions of the meta-feature embeddings from $\{1,2,4,8,16\}$.
See text for discussion.
}
\label{table:space-efficient-vis-rec-variants-vary-MFE-dimension}
\vspace{-2mm}
\footnotesize
\begin{tabular}{l c ccccc   ccccc   }
\toprule
 &&\multicolumn{5}{c}{HR@K} & \multicolumn{5}{c}{NDCG@K} \\ 
\cmidrule(lr){3-7} 
\cmidrule(lr){8-12} 
 \textbf{Model} & \textbf{MFE dim.} & @1 & @2 & @3 & @4 & @5 & @1 & @2 & @3 & @4 & @5 \\
 \midrule
  PVisRec ($\mA$,$\mC$,$\mM$ only) & 1 & 0.284 & 0.413 & 0.480 & 0.512 & 0.529 & 0.282 & 0.364 & 0.398 & 0.412 & 0.418 \\
& 2 &  0.245 & 0.348 & 0.395 & 0.417 & 0.429 & 0.244 & 0.308 & 0.333 & 0.342 & 0.346 \\
& 4 & 0.265 & 0.388 & 0.444 & 0.468 & 0.481 & 0.263 & 0.341 & 0.369 & 0.380 & 0.385 \\
& 8 &  0.304 & 0.419 & 0.462 & 0.492 & 0.506 & 0.302 & 0.376 & 0.397 & 0.410 & 0.416 \\
& 16 & 0.294 & 0.404 & 0.452 & 0.471 & 0.483 & 0.292 & 0.362 & 0.386 & 0.395 & 0.399 \\
\midrule
  PVisRec & 1 &  0.467 & 0.589 & 0.641 & 0.667 & 0.681 & 0.464 & 0.542 & 0.569 & 0.580 & 0.585 \\
& 2 & 0.542 & 0.685 & 0.744 & 0.771 & 0.792 & 0.539 & 0.630 & 0.660 & 0.672 & 0.680 \\
& 4 & 0.544 & 0.713 & 0.779 & 0.815 & 0.829 & 0.541 & 0.649 & 0.682 & 0.698 & 0.704 \\
& 8 &  0.608 & 0.806 & 0.874 & 0.906 & 0.925 & 0.604 & 0.731 & 0.765 & 0.779 & 0.787 \\
& 16 & 0.616 & 0.794 & 0.865 & 0.896 & 0.916 & 0.613 & 0.726 & 0.762 & 0.776 & 0.784 \\
\bottomrule
\end{tabular}
\end{table}

\subsection{Improving Space-Efficiency via Meta-Feature Embeddings} 
\label{sec:exp-space-efficient-results-meta-feature-embeddings}
In this section, we investigate using a low-rank meta-feature embedding matrix to significantly improve the space-efficiency of our proposed approach.
In particular, we replace the original meta-feature matrix $\mM$ with a low-rank approximation that captures the most important and meaningful meta-feature signals in the data. 
In addition to significantly reducing the space requirements of PVisRec, we also investigate the performance when the low-rank meta-feature embeddings are used, and the space and accuracy trade-off as the number of meta-feature embedding dimensions varies from $\{1,2,4,8,16\}$.
We set $d=10$ and vary the dimensions of the dimensionality of the meta-feature embeddings (MFE) from $\{1,2,4,8,16\}$ across the different proposed approaches.
We provide the results in Table~\ref{table:space-efficient-vis-rec-variants-vary-MFE-dimension} for the space-efficient variants of our personalized visualization recommendation methods that use meta-feature embeddings.
Overall, we find that in nearly all cases, we find similar HR@K and NDCG@K compared to the original variants, while obtaining a significantly more compact model with orders of magnitude less space.
For instance, when MFE dim. is 16, PvisRec has a HIT@1 of 0.616 compared to 0.630 using the original 1006-dimensional meta-feature matrix, which uses roughly 63x more space compared to the 16-dimensional MFE variant.
As an aside, since PVisRec ($\mA$, $\mC$, $\mD$) does not use $\mM$, it does not have a meta-feature embedding (MFE) variant.
This implies that we can indeed significantly reduce the space requirements of our approaches by trading off only a tiny amount of accuracy \textcolor{black}{\textbf{(RQ3)}}.

\subsection{Neural Personalized Visualization Recommendation}
\label{sec:exp-adding-DL-component}
In this section, we study the performance of the proposed Neural Personalized Visualization Recommendation models (\textbf{RQ4}).
For these experiments, we use $d=10$ and $\alpha=0.5$ for Neural PVisRec-CMF.
All models use three layers and ReLU for the activation function.
See Section~\ref{sec:neural-pvisrec} for further details.
The results are provided in Table~\ref{table:vis-rec-neural-variants}.
Both neural personalized visualization recommendation models outperform the simpler and faster graph-based PVisRec approach (across both rank-based evaluation metrics and across all top-$K$ personalized visualization recommendations).
This is expected since the neural visualization models all leverage the graph-based PVisRec model in some fashion.
Neural PVisRec uses the learned low-dimensional embeddings of the users, visual-configurations, attributes, and meta-features of the attributes as input into the first layer whereas Neural PVisRec-CMF also uses the learned low-dimensional embeddings, but also uses the predicted visualization scores from the PVisRec model for each user and combines these with the predicted scores from the neural component.
Notably, both neural personalized visualization recommendation models outperform the simpler and faster graph-based approach.
In Table~\ref{table:vis-rec-neural-variants}, Neural PVisRec-CMF outperforms the simpler Neural PVisRec network. This holds for HR@$K$ and NDCG@$K$, and across all top-$K$ personalized visualization recommendations where $K\in \{1,...,5\}$.

\begin{table}[h!]
\caption{Results for the \emph{Neural} Personalized Visualization Recommendation Models.}
\label{table:vis-rec-neural-variants}
\vspace{-2mm}
\footnotesize
\begin{tabular}{l ccccc   ccccc   }
\toprule
&\multicolumn{5}{c}{HR@K} &\multicolumn{5}{c}{NDCG@K} \\
\cmidrule(lr){2-6} 
\cmidrule(lr){7-11} 
\textbf{Model} & @1 & @2 & @3 & @4 & @5 & @1 & @2 & @3 & @4 & @5 \\
\midrule
 Neural PVisRec & 0.656 & 0.825 & 0.889 & 0.923 & 0.946 & 0.652 & 0.761 & 0.793 & 0.808 & 0.817 \\
 Neural PVisRec-CMF & 0.762 & 0.879 & 0.922 & 0.944 & 0.961 & 
0.729 & 0.822 & 0.845 & 0.855 & 0.861 \\
\bottomrule
\end{tabular}
\end{table}

\subsubsection{Nonlinear activation function.}
Neural PVisRec is flexible and can leverage any nonlinear activation functions for the fully-connected layers of our multilayer neural network architecture for personalized visualization recommendation.
In Table~\ref{table:vis-rec-variants-ablation-vary-activation-func}, we compare three non-linear activation functions $\sigma$ for learning a personalized visualization recommendation model including 
hyperbolic tangent (tanh) $\sigma(\vx)=\tanh(\vx)$, 
sigmoid $\sigma(\vx) = 1 / (1+\exp[-\vx])$, and
ReLU $\sigma(\vx)=\max(0,\vx)$.
The results in Table~\ref{table:vis-rec-variants-ablation-vary-activation-func} show that ReLU performs best by a large margin followed by sigmoid and then tanh.
ReLU likely performs well due to its ability to avoid saturation, handle sparse data and be less likely to overfit.

\begin{table}[h!]
\caption{Ablation study results of Neural PVisRec with different nonlinear activation functions.
We report HR@1 for brevity.
All results use $d=10$ and $\alpha=0.5$.
}
\label{table:vis-rec-variants-ablation-vary-activation-func}
\vspace{-2mm}
\footnotesize
\begin{tabular}{l  ccccc}
\toprule
&\multicolumn{3}{c}{nonlinear activation $\sigma$} & \\ 
\cmidrule(lr){2-5} 
\textbf{Model} & tanh & sigmoid &  ReLU & \\
\midrule
 Neural PVisRec & 0.615 & 0.624 & 0.656 & \\
 Neural PVisRec-CMF & 0.613 & 0.640 & 0.762 & \\
\bottomrule
\end{tabular}
\end{table}

\begin{table}[b!]
\caption{Comparing performance of Neural PVisRec with different number of hidden layers.
}
\label{table:vis-rec-variants-ablation-vary-num-hidden-layers}
\vspace{-2mm}
\footnotesize
\begin{tabular}{c ccccc   ccccc}
\toprule
& \multicolumn{5}{c}{HR@K} & \multicolumn{5}{c}{NDCG@K} \\
\cmidrule(lr){2-6} 
\cmidrule(lr){7-11} 
\textbf{\# Hidden Layers} & @1 & @2 & @3 & @4 & @5 & @1 & @2 & @3 & @4 & @5 \\
\midrule
1 & 0.579 & 0.773 & 0.844 & 0.880 & 0.896 & 0.578 & 0.701 & 0.737 & 0.752 & 0.758 \\
2 & 0.618 & 0.801 & 0.865 & 0.892 & 0.907 & 0.618 & 0.733 & 0.765 & 0.777 & 0.783 \\
3 & 0.656 & 0.825 & 0.889 & 0.923 & 0.946 & 0.652 & 0.761 & 0.793 & 0.808 & 0.817 \\
4 & 0.646 & 0.754 & 0.813 & 0.842 & 0.869 & 0.499 & 0.639 & 0.680 & 0.694 & 0.705 \\
\bottomrule
\end{tabular}
\end{table}

\subsubsection{Hidden layers.}
To understand the impact of the number of layers on the performance of the neural personalized visualization recommendation models, we vary the number of hidden layers from $L\in\{1,2,3,4\}$.
In Table~\ref{table:vis-rec-variants-ablation-vary-num-hidden-layers}, the performance increases as additional hidden layers are included, and begins to decrease at $L=4$.
The best performance is achieved with three hidden layers.
This result indicates the benefit of deep learning for personalized visualization recommendation.

\subsubsection{Layer size.}
Recall that our network structure followed a tower pattern where the layer size of each successive layer is halved.
In this experiment, we investigate larger layer sizes while fixing the final output embedding size to be 8 and using 4 hidden layers.
In Table~\ref{table:vis-rec-neural-variants-layer-sizes-tower-fac-architecture}, we observe a significant improvement in the visualization ranking when using larger layer sizes. 

\begin{table}[h!]
\caption{Varying layer sizes in the deep personalized visualization recommendation model (Neural PVisRec).}
\label{table:vis-rec-neural-variants-layer-sizes-tower-fac-architecture}
\vspace{-2mm}
\footnotesize
\begin{tabular}{l ccccc   }
\toprule
& \multicolumn{5}{c}{HR@K}  \\ 
\cmidrule(lr){2-6} 
 \textbf{Layer Sizes} & @1 & @2 & @3 & @4 & @5 \\
 \midrule
 8-16-32-64 & 0.701 & 0.790 & 0.832 & 0.865 & 0.883 \\
 8-32-128-512 & 0.734 & 0.797 & 0.846 & 0.874 & 0.886 \\
 8-48-288-1728 & 0.752 & 0.822 & 0.869 & 0.895 & 0.913 \\
\bottomrule
\end{tabular}
\end{table}

\subsubsection{Runtime performance.}
Neural PVisRec is also fast, taking on average 10.85 seconds to train using the large personalized visualization corpus from Section~\ref{sec:dataset}.
The other neural visualization recommender is nearly as fast, as it contains only an additional step that is linear in the output embedding size.
For these experiments, we used a 2017 MacBook Pro with 16GB memory and 3.1GHz Intel Core i7 processor.

\section{Related Work}\label{sec:related-work}

\subsection{Visualization Recommendation}
Rule-based visualization recommendation systems such as Voyager~\cite{vartak2017towards, voyager2, voyager}, VizDeck~\cite{perry2013vizdeck}, and DIVE~\cite{hu2018dive} use a large set of rules  defined manually by domain experts to recommend appropriate visualizations that satisfy the rules~\cite{dorisjang,mackinlay1986automating,roth1994interactive,casner1991taskBOZ,mackinlay2007show, derthick1997interactive, stolte2002polaris,feiner1985apex,seo2005rankbyfeature}.
Such rule-based systems do not leverage any training data for learning or user personalization. 
There have been a few ``hybrid'' approaches that combine some form of learning with manually defined rules for visualization recommendation~\cite{draco}, \eg, Draco learns weights for rules (constraints)~\cite{draco}.
Recently, there has been work that focused on the end-to-end ML-based visualization recommendation problem~\cite{ML-based-Vis-Rec,data2vis}.
However, this work learns a global visualization recommendation model that is agnostic of the user, and thus not able to be used for the personalized visualization recommendation problem studied in our work.

All of the existing rule-based~\cite{vartak2017towards, voyager2, voyager, perry2013vizdeck, hu2018dive}, hybrid~\cite{draco}, and pure ML-based visualization recommendation~\cite{ML-based-Vis-Rec} approaches are unable to recommend personalized visualizations for specific users.
These approaches do not model users, but focus entirely on learning or manually defining visualization rules that capture the notion of an effective visualization~\cite{mackinlay2007showme,wills2010autovis,key2012vizdeck,van2013small,
wilkinson2008scagnostics,dang2014scagexplorer,vartak2015seedb, demiralp2017foresight-vldb,cui2019datasite,lin2020dziban,lee2019vispilot,siddiqui2016zenvisage}.
Therefore, no matter the user, the model always gives the same recommendations.
The closest existing work is VizRec~\cite{vizrec}.
However, VizRec is only applicable when there is a single dataset shared by all users (and therefore a single small set of visualizations 
that the users have explicitly liked and tagged).
This problem is unrealistic with many impractical assumptions that are not aligned with practice.
Nevertheless, the problem solved by that prior work is
a simple special case of the personalized visualization recommendation problem introduced in our paper.

\subsection{Simpler Design and Data Tasks}
Besides visualization recommendation, there are methods that solve simpler sub-tasks such as improving expressiveness, improving perceptual effectiveness, matching task types, etc. 
These simpler sub-tasks can generally be divided two categories~\cite{dorisjang, wongsuphasawat2016towards}: 
whether the solution focuses on recommending data (\textit{what data to visualize}), such as Discovery-driven Data Cubes~\cite{sarawagi1998discovery}, Scagnostics~\cite{wilkinson2005graph}, AutoVis~\cite{wills2010autovis}, and MuVE~\cite{ehsan2016muve}) or recommending encoding (\textit{how to design and visually encode the data}), such as APT~\cite{mackinlay1986automating}, ShowMe~\cite{mackinlay2007show}, and Draco--learn~\cite{draco}). 
While some of those are ML-based, none are able to recommend entire visualizations (nor are they personalized), which is the focus of this work.
For example, VizML~\cite{vizml} 
predicts the type of a chart (e.g., bar, scatter, etc.) instead of complete visualization.
Draco~\cite{draco} infers weights for a set of manually defined rules.
VisPilot~\cite{lee2019avoiding} recommended different drill-down data subsets from datasets. 
As an aside, not only does these works not solve the visualization recommendation problem, they are also not personalized for individual users.
Instead of solving simple sub-tasks such as predicting the chart type of a visualization, 
we focus on the \emph{end-to-end personalized visualization recommendation problem (Sec.~\ref{sec:problem})}:
given a dataset of interest to user $i$,
the goal is to \textit{automatically recommend the top-k most effective visualizations personalized for that individual user.}
This paper fills the gap by proposing the first \emph{personalized} visualization recommendation approach that is completely automatic, data-driven, 
and most importantly recommends personalized visualizations based on a users previous feedback, behavior, and interactions with the system.

\subsection{Traditional Recommender Systems}
In traditional item-based recommender systems~\cite{adomavicius2005toward,ricci2011introduction,zhao2020catn-cross-domain,noel2012new,zhang2017joint}, there is a \emph{single shared set of items} (\ie, movies~\cite{bennett2007netflix, covington2016deep, harper2015movielens}, products~\cite{linden2003amazon}, hashtags~\cite{sigurbjornsson2008flickr, wang2020earn}, documents~\cite{xu2020understanding, kanakia2019scalable},  news~\cite{ge2020graph}, books~\cite{liu2014recommending}, and location~\cite{Mao11exploiting, zhou2019adversarial,bennett2011inferring}).
However, in the personalized visualization recommendation problem studied in this work,
since visualizations are dataset dependent, there is not a shared set of visualizations to recommend users.
Therefore, given $N$ datasets, there are $N$ completely disjoint sets of visualizations that can be recommended.
Every dataset consists of its own completely separate set of relevant visualizations that are exclusive to the dataset. 
Therefore, in contrast to the goal of traditional item recommender systems, 
the goal of personalized visualization recommendation is to learn a personalized vis. rec. model for each individual user, which is capable of scoring and ultimately recommending personalized visualizations to that user from any unseen dataset in the future.
Some recent works have adapted various deep learning approaches for collaborative filtering~\cite{sedhain2015autorec,li2020learning,ncf,chen2020efficient,guan2019attentive}.
However, none of these works have focused on the problem of \emph{personalized visualization recommendation} studied in this work.
The personalized visualization recommender problem has a few similarities with cross-domain recommendation~\cite{tang2012cross,gao2013cross,man2017cross,shapira2013facebook,hu2013personalized}.
In cross-domain item recommendation, there is only a few datasets as opposed to tens of thousands of different datasets in our problem~\cite{zhao2020catn-cross-domain}.
More importantly, in cross-domain item recommendation, the different datasets are assumed to share at least one mode between each other, whereas in personalized visualization recommendation, each new dataset gives rise to a completely different set of visualizations to recommend.

\section{Conclusion} \label{sec:conc}
In this work, we introduced the problem of user-specific personalized visualization recommendation and proposed an approach for solving it.
The approach learns individual personalized visualization recommendation models for each user.
In particular, the personalized vis. rec. models for each user are learned by taking into account the user feedback including both implicit and explicit feedback regarding the visual and data preferences of the users, as well as users whom have also explored similar datasets and visualizations.
We overcome the issues with data sparsity and limited user feedback by leveraging the data and visualization preferences of users whom are similar, despite that the visualizations from such users are from completely different datasets.
The models are able to learn better visualization recommendation models for each individual user by leveraging the data and visualization preferences of users whom are similar.
In addition, we proposed a deep neural network architecture for \emph{neural} personalized visualization recommendation that can learn complex non-linear relationships between the users, their attributes of interest, and visualization preferences.
This paper is a first step in the direction of learning personalized visualization recommendation models for individual users based on their data and visualization feedback, and the data and visual preferences of users with similar data and visual preferences.
Future work should investigate and develop better machine learning models and learning techniques to further improve the personalized visualization recommendation models and the visualization recommendations for individual users.

\balance
\bibliographystyle{ACM-Reference-Format}
\bibliography{paper}

\end{document}